\documentstyle[12pt,epsf]{article}
\baselineskip=\normalbaselineskip
\ifx\TwoupWrites\UnDeFiNeD\else\target{\magstepminus1}{11.3in}{8.27in}
	\source{\magstep0}{7.5in}{11.69in}\fi
\newfont{\fourteencp}{cmcsc10 scaled\magstep2}
\newfont{\titlefont}{cmbx10 scaled\magstep2}
\newfont{\authorfont}{cmcsc10 scaled\magstep1}
\newfont{\fourteenmib}{cmmib10 scaled\magstep2}
	\skewchar\fourteenmib='177
\newfont{\elevenmib}{cmmib10 scaled\magstephalf}
	\skewchar\elevenmib='177
\makeatletter
\newcommand\nonsequentialeqnum{
	\@addtoreset{equation}{section}
	\def\theequation{\arabic{section}.\arabic{equation}}}
\newif\ifp@bblock  \p@bblocktrue
\newcommand\nopubblock{\p@bblockfalse}
\newcommand\topspace{\hrule height 0pt depth 0pt \vskip}
\newcommand\p@bblock{\begingroup \tabskip=\hsize minus \hsize
	\baselineskip=1.5\ht\strutbox \topspace-2\baselineskip
	\halign to\hsize{\strut ##\hfil\tabskip=0pt\crcr
	\the\Pubnum\crcr
	\the\date\crcr}\endgroup}
\newcommand\YITPmark{\hbox{\fourteenmib YITP\hskip0.2cm
        \elevenmib Uji\hskip0.15cm Research\hskip0.15cm Center\hfill}}
\renewcommand\titlepage{\ifx\TwoupWrites\UnDeFiNeD\null\vspace{-1.7cm}\fi
	\YITPmark\vskip0.6cm
	\ifp@bblock\p@bblock \else\hrule height 0pt \relax \fi}
\makeatother
\newtoks\date
\newtoks\Pubnum
\newtoks\pubnum
\Pubnum={YITP/U-\the\pubnum\cr gr-qc/9507037}
\date={\today}
\newcommand{\frontpageskip}{\vspace{12pt plus .5fil minus 2pt}}
\renewcommand{\title}[1]{\frontpageskip
	\begin{center}{\titlefont #1}\end{center}\par}
\renewcommand{\author}[1]{\frontpageskip\par\begin{center}
	{\authorfont #1}\end{center}
	\nobreak
	}

\newcommand{\address}[1]{\par\begin{center}{\sl #1}\end{center}\par}

\renewcommand{\thanks}[1]{\footnote{#1}}
\renewcommand{\abstract}{\par\frontpageskip\centerline{\fourteencp Abstract}
	\vspace{8pt plus 3pt minus 3pt}}

\catcode`@=11
\@addtoreset{equation}{section}   % Makes \section reset 'equation' counter.
\def\theequation{\arabic{section}.\arabic{equation}}
\def\thebibliography#1{\section*{References\@mkboth
 {REFERENCES}{REFERENCES}}\list
 {\leftbibmark\arabic{enumi}\rightbibmark}{
 \settowidth\labelwidth{\leftbibmark #1\rightbibmark}\leftmargin\labelwidth
 \advance\leftmargin\labelsep
 \usecounter{enumi}}
 \def\newblock{\hskip .11em plus .33em minus -.07em}
 \sloppy\clubpenalty4000\widowpenalty4000
 \sfcode`\.=1000\relax}

\def\@cite#1#2{\leftcitemark{#1\if@tempswa , #2\fi}\rightcitemark}
\def\leftcitemark{[}
\def\rightcitemark{]}
\def\leftbibmark{[}
\def\rightbibmark{]}
\def\CITE#1{$^{\hbox{\small \cite{#1}}}$}

\catcode`@=10
\def\getlength#1{\ifx#1\end \let\next=\relax
    \else\advance\count0 by1 \let\next=\getlength \fi \next}
\def\emptybox#1#2{\framebox[#1]{\rule{0cm}{#2}}}   % #1:width #2 height
\def\LeftFigure#1#2#3#4#5{
\begin{figure}
\noindent\emptybox{#1}{#2}\hfill
{\addtolength{\hsize}{-#1}\addtolength{\hsize}{-1cm}
\parbox[b]{\hsize}{\caption{\label{#3}#4}
{\count0=0 \getlength #5\end \ifnum\count0>0 \hfill\\ {\small #5} \fi}\par}
}
\end{figure}
}
\def\Figure#1#2#3#4#5{
\begin{figure}
\emptybox{#1}{#2}
\caption{\label{#3}#4} {\count0=0 \getlength #5\end \ifnum\count0>0 \hfill\\
{\small #5} \fi}
\end{figure}
}

\def\A{{\cal A}}
\def\B{{\cal B}}
\def\C{{\cal C}}
\def\D{{\cal D}}

\def\H{{\cal H}}
\def\I{{\cal I}}
\def\J{{\cal J}}
\def\K{{\cal K}}
\def\L{{\cal L}}
\def\M{{\cal M}}
\def\N{{\cal N}}

\def\SS{{\cal S}}

\def\U{{\cal U}}
\def\V{{\cal V}}
\def\W{{\cal W}}

\def\Z{{\cal Z}}

\def\linebreak{\hfill\break}

\def\newterm#1{{\it #1}}
\def\Eq#1{Eq.(\ref{#1})}
\def\Eqs#1#2{Eqs.(\ref{#1})-(\ref{#2})}
\def\Prop#1{Proposition \ref{#1}}
\def\Cor#1{Corollary \ref{#1}}

\def\Theorem#1{Theorem \ref{#1}}

\def\tend{\rightarrow}

\def\equivalent{\quad\Leftrightarrow\quad}
\def\therefore{\mbox{\setbox0=\hbox{X}\hbox{$\ldotp$}\raise0.7\ht0\hbox{$\ldotp$}\hbox{$\ldotp$}} }
\def\because{\mbox{\setbox0=\hbox{X}\raise0.7\ht0\hbox{$\ldotp$}\hbox{$\ldotp$}\raise0.7\ht0\hbox{$\ldotp$}}\kern0pt }
\def\r#1{{\rm #1}}

\def\bm#1{\mbox{\boldmath $#1$}}
\let\bg=\bm
\def\Frac(#1/#2){\left(\frac{#1}{#2}\right)}
\def\Tr{\r{Tr}}

\def\RF{{\bm{R}}}
\def\CF{{\bm{C}}}

\def\sRF{{\bm{\scriptstyle R}}}

\def\orth{\perp}

\def\Hom(#1){{\rm Hom}(#1)}
\def\range#1{{\rm ran}\; #1}
\def\domain#1{{\rm dom}\; #1}
\def\kernel#1{{\rm ker}\; #1}

\def\spectrum#1{{\rm spec}(#1)}
\def\VNA{W^*}

\def\SOT{{\rm SOT}}

\def\maps{\rightarrow}

\def\In{\mathrel{\mbox{\setbox0=\hbox{$\cup$}\dimen0=\wd0\divide\dimen0 by 2
\box0\kern -\dimen0\vrule}}}
\def\Mapping#1#2#3#4{\begin{array}[t]{ccc}
#1 & \rightarrow & #3\\
\In &            & \In\\
#2 &             & #4 \end{array}}

\def\SetDef#1#2{\left\{#1\;|\;#2\right\}}

\newtheorem{theorem}{Theorem}[section]
\newtheorem{proposition}{Proposition}[section]
\newtheorem{corollary}{Corollary}[section]
\newtheorem{lemma}{Lemma}[section]

\def\FilledSquare{\vbox{\setbox1=\hbox{X} \hrule height\ht1 width0.5\wd1
depth0pt}}
\newenvironment{proof}{\par\noindent {\bf Proof}\par}{\par\medskip}
\def\QED{ \FilledSquare}

\renewenvironment{matrix}[1]{\left(\begin{array}{#1}}{\end{array}\right)}

\def\Beq{\begin{equation}}
\def\Eeq{\end{equation}}
\def\Beqr{\begin{eqnarray}}
\def\Eeqr{\end{eqnarray}}
\def\Beqrn{\begin{eqnarray*}}
\def\Eeqrn{\end{eqnarray*}}
\def\Bitm{\begin{itemize}}
\def\Eitm{\end{itemize}}

\begin{document}
\thispagestyle{empty}
%
%\nopubblock        %% uncomment in making submit-version
%\nonsequentialeqnum %% uncomment in (Section.Number) equation number style.
\pubnum{95-21}
\date{July 1995}
\titlepage

\title{
Dynamics of Totally Constrained Systems \\
II. Quantum Theory
}

\author{
Hideo Kodama
}
\address{
Yukawa Institute for Theoretical Physics, Kyoto University,
Uji 611, Japan
}

\abstract{
In this paper a new formulation of quantum dynamics of totally
constrained systems is developed, in which physical quantities
representing time are included as observables. In this formulation
the hamiltonian constraints are imposed on a relative probability amplitude
functional $\Psi$ which determines the relative probability for each
state to be observed, instead of on the state vectors as in the
conventional Dirac quantization.  This leads to a foliation of
the state space by linear manifolds on each of which $\Psi$ is
constant, and dynamics is described as linear mappings among
acausal subspaces which are transversal to these linear manifolds.
This is a quantum analogue of the classical statistical dynamics
of totally constrained systems developed in the previous paper.
It is shown that if the von Neumann algebra $\C$ generated by the constant
of motion is of type I, $\Psi$ can be consistently normalizable on
the acausal subspaces on which a factor subalgebra of $\C$ is
represented irreducibly, and the mappings among these acausal
subspaces are conformal. How the formulation works is illustrated by
simple totally constrained systems with a single constraint such as
the parametrized quantum mechanics, a relativistic free particle in
Minkowski and curved spacetimes, and a simple minisuperspace model.
It is pointed out that the inner product of the relative probability
amplitudes induced from the original Hilbert space picks up a special
decomposition of the wave functions to the positive and the negative
frequency modes.
}

\newpage

\section{Introduction}

As shown in the previous paper\CITE{Kodama.H1995}, in a totally
constrained system $(\Gamma,\omega,\{h_\alpha\})$, the involutive
system of the infinitesimal canonical transformations $Y_\alpha$
generated by the constraint functions $h_\alpha$ defines a foliation
of a neighborhood of the constraint submanifold $\Sigma_H$, and the
dynamics of an ensemble of the system is described by the causal
mapping among acausal submanifolds which are transversal to the
foliation.  In particular the statistical dynamics of the system
can be consistently formulated in terms of a relative distribution
function $\rho$ which is preserved by the causal mapping.

In this paper we develop a new formulation of quantum dynamics of
totally constrained systems by introducing a structure to a
state space analogous to this dynamical structure of the classical
system.  The key idea is to adopt a probability amplitude functional
$\Psi$ on the state space, which corresponds to the relative distribution
function $\rho$ in the classical statistical dynamics and is not bounded
in general, to describe a dynamical state, and to impose the
hamiltonian constraints on this functional.  This is based on the
philosophy that the hamiltonian  constraints should be imposed
on a dynamical object which selects possible state vectors instead of
on the state vectors themselves\CITE{Kodama.H1995,Kodama.H1993}.

In this formulation the quantum dynamics is described by a linear
mapping between two acausal subspaces which are transversal to
the foliation of the state space.  Thus it has a similar structure
to that of the classical theory, but various new problems arise in the
quantum theory.  In particular the linear mappings among acausal
subspaces are not unitary in general.  To give a prescription to
resolve this difficulty consists the main part of this paper.
The basic idea is to restrict the acausal subspaces to those
which are invariant under the operation of  a physical subalgebra
of the von Neumann algebra generated by constants of motion.

The organization of the paper is as follows. First in the next section
on the basis of the analogy with the dynamics of classical totally
constrained systems the hamiltonian constraint is formulated as a
constraint on the relative probability amplitude, which is called the
weak hamiltonian constraint, and it is outlined how to describe
quantum dynamics in terms of the relative probability amplitude
satisfying the weak hamiltonian constraint. Further it is pointed out
that the unitarity of dynamics is violated in this formulation unless
some additional restriction on the choice of physical acausal
subspaces, which play the role of the state space in the ordinary
quantum mechanics, is introduced. Then in \S3 the concepts of physical
subalgebras and physical acausal subspaces are introduced with the
help of the central decomposition of the von Neumann algebra $\C$
formed by constants of motion, and it is shown that the unitarity
problem is resolved by restricting the acausal subspaces to those on
which a physical subalgebra is irreducibly represented if $\C$ is of
type I. In \S4 this abstract formulation is applied to simple totally
constrained systems with a single constraint in order to illustrate
how it works. As a byproduct it is shown that for the system of a free
relativistic particle in curved spacetime the inner product of
relative probability amplitudes induced from the inner product of the
original Hilbert space picks up a special decomposition into the
positive and negative frequency parts of the relative probability
amplitudes, which correspond to the ordinary wave function satisfying
the Wheeler-DeWitt equation in the minisuperspace models. Section 5 is
devoted to some discussions.

\section{Formulation}

In this paper we start from a mathematically well-defined pair of a
Hilbert space $\H$ and a set of operators $\{ h_\alpha \}$ on $\H$
which are obtained by quantization of a classical totally constrained
system $(\Gamma,\omega,\{h_\alpha\})$, and call the pair $(\H, \{h_\alpha\})$
a quantum totally constrained system. We do not discuss the details of
the quantization procedure or the problems associated with it, though
some comments are given in \S4.4. Throughout the paper we assume that
$\H$ is separable and the constraint operators are self-adjoint.
For notational simplicity we use the same symbols for functions on
the classical phase space $\Gamma$ and the corresponding operators
on $\H$ as far as no confusion occurs.

\subsection{Quantum hamiltonian constraint}

As explained in the previous paper\CITE{Kodama.H1995}, the
infinitesimal canonical transformations $Y_\alpha$ generated
by the constraint functions $h_\alpha$ give an involutive system
on the constraint submanifold $\Sigma_H$ and the causal submanifolds
defined as connected components of its integration submanifolds
define a foliation of a neighborhood of $\Sigma_H$ in the phase
space $\Gamma$. In this foliation each causal submanifold is
one-to-one correspondence to a dynamical state of the single system, while
each point in $\Gamma$ represents a possible state of the system at some
instant. Further if we consider an ensemble of the system, its
dynamical state is described by a relative distribution function
$\rho$ on $\Gamma$ which has its support in $\Sigma_H$ and is
constant on each causal submanifold.  These conditions on $\rho$
are expressed by the dynamical equations
\Beqr
&&h_\alpha\rho=0,\label{EqForRho1}\\
&&Y_\alpha\rho=0.\label{EqForRho2}
\Eeqr

When we turn to the quantum system, the classical phase space $\Gamma$
is replaced by the Hilbert space $\H$.  Hence, taking account of
the linear superposition principle of quantum theory, it is natural
to consider a linear functional $\Psi$ on $\H$ such that the
relative probability for each state $u\in\H$ to be observed is given by
\Beq
\Pr(u) \propto |\Psi(u)|^2,
\Eeq
as an object to describe the dynamical state of the system corresponding
to $\rho$ in the classical theory.  Let us call such this functional
\newterm{the relative probability amplitude}.

If we push this correspondence further, the infinitesimal canonical
transformation $\sum_\alpha \lambda^\alpha Y_\alpha$(finite sum;
$\lambda^a\in\RF$)
corresponds the self-adjoint operator $\sum_\alpha \lambda^\alpha h_\alpha$
which generate a unitary transformation $\exp(i\sum_\alpha\lambda^\alpha
h_\alpha)$ on $\H$. Hence it  is natural to replace the requirement
that $\rho$ is constant along each causal submanifold by
the requirement
\Beq
\Psi(e^{i\sum_\alpha\lambda^\alpha h_\alpha}u)=\Psi(u)
\quad \forall \lambda^\alpha, \quad \forall u\in\H,
\Eeq
or its infinitesimal form
\Beq
\Psi(h_\alpha u)=0 \quad \forall \alpha, \quad
\forall u\in\domain{h_\alpha}.
\label{WeakHamiltonianConstraint}
\Eeq

To be exact, these two equations are not equivalent and the former
is stronger than the latter in general because we do not require the
boundedness of $\Psi$ as will be explained below. In the present paper
we adopt the latter one as the quantum expression of the classical
hamiltonian constraint, and call it the \newterm{weak hamiltonian constraint}
since it has a better correspondence with the conventional Dirac
constraint\CITE{Dirac.P1964B}
as well as with the classical conditions \Eqs{EqForRho1}{EqForRho2}.

If we define the linear submanifold $\N$ of $\H$ by
\Beq
\N=\sum_\alpha \range{h_\alpha} \quad (\r{finite\ sum}),
\label{NullSpace:def}\Eeq
the weak hamiltonian constraint is expressed as
\Beq
\Psi(\N)=0 \equivalent \Psi(u)=0 \quad\forall u\in\N.
\Eeq
Hence if we foliate the Hilbert space $\H$ by $\N$ and its translations
as
\Beq
\H=\bigcup_\lambda \N_\lambda; \quad
u,v\in\N_\lambda \equivalent u-v\in\N,
\Eeq
the weak hamiltonian constraint implies that $\Psi$ is constant on
each leaf $\N_\lambda$, i.e.,
\Beq
\Psi(u+\N)=\Psi(u) \quad \forall u\in\N.
\Eeq
We call this linear manifold $\N$ the \newterm{linear null submanifold}.

Here note that if $\Psi$ is bounded or equivalently continuous on $\H$,
there exists a unique state vector $\Phi\in\H$ such that
$\Psi(u)=(\Phi,u)$ from the Riesz theorem, and the weak hamiltonian
constraints coincide with the constraints $h_\alpha\Phi=0$ in the
standard Dirac quantization. This implies that all the difficulties of
the Dirac quantization remain.  Therefore we do not require that $\Psi$
is a bounded functional on $\H$.  This is consistent with the fact that
$\Psi$ corresponds to $\rho$ which is not normalizable.

The argument so far is concerned with the description
of the quantum system in a pure state. It is easy to extend
the description to mixed states. For that purpose let us introduce
a bilinear functional $R$ on $\H$ which gives the probability
for a state $u$ to be observed by
\Beq
\Pr(u) \propto R(u,u).
\Eeq
As in the case of $\Psi$, we do not require that $R$ is continuous
on $\H$, but we require that it satisfies
\Beqr
&&R(u,a_1v_1+a_2v_2)=a_1R(u,v_1)+a_2R(u,v_2),\\
&&R(u,v)=\overline{R(v,u)},
\Eeqr
in order to guarantee the positivity of $R(u,u)$ and
to respect the superposition principle. For a pure dynamical state
represented by $\Psi$ we assume that the corresponding $R$ is
given by
\Beq
R(u,v)=\overline{\Psi(u)}\Psi(v).
\Eeq

The same argument as for $\Psi$ yields the following equation on $R$:
\Beq
R(u,h_\alpha v)=R(h_\alpha u, v) \quad \forall u,v\in\domain{h_\alpha}.
\label{WWeakHamiltonianConstraint}\Eeq
However, this equation does not reduce to the weak hamiltonian constraint
on $\Psi$ when $R$ corresponds to a pure state given by $\Psi$.
Instead it yields a weaker equation
\Beq
\Psi((h_\alpha-c_\alpha)u)=0,
\Eeq
where $c_\alpha$ are some real constants. Obviously in order to
get the weak hamiltonian constraints, we must replace the above equation
by the stronger one,
\Beq
R(u, h_\alpha v)=0 \quad \forall u\in\H, \quad
\forall v\in\domain{h_\alpha}.
\Eeq
\label{WeakHamiltonianConstraintOnRho}
We call this equation the weak hamiltonian constraint on
the \newterm{relative density functional} $R$.

The reason why we must impose a stronger constraint on $R$ is related
to the correspondence between the quantum constraint and the
classical constraint. To see this, let us forget about the mathematical
rigor for a while, and assume that $R$ can be expressed formally in
terms of a self-adjoint operator $\rho$ as
\Beq
R(u,v)=(u,\rho v).
\Eeq
Then the weaker equation (\ref{WWeakHamiltonianConstraint}) is written as
\Beq
[\rho, h]=0, \label{tmp:wwconstraint}
\Eeq
which corresponds to \Eq{EqForRho2} in the classical limit.  On the other
hand \Eq{WeakHamiltonianConstraintOnRho} yields
\Beq
\rho h+ h\rho=0
\Eeq
in addition to \Eq{tmp:wwconstraint}, which reduces to \Eq{EqForRho2} in the
classical limit.  Hence we can expect that the classical hamiltonian
constraints are recovered in the classical limit only when we impose
the stronger equation \Eq{WeakHamiltonianConstraintOnRho} on $R$.

On the basis of these heuristic arguments we postulate that
the classical constraints are replaced by the weak hamiltonian
constraints \Eq{WeakHamiltonianConstraint} or
\Eq{WeakHamiltonianConstraintOnRho} in quantum theory. Note
that, exactly speaking, these quantum constraints contain the
dynamical equation in addition to the constraint because they correspond
to the two equations (\ref{EqForRho1}) and (\ref{EqForRho2}) in the
classical theory.

\subsection{Quantum Dynamics}

The relative probability amplitude $\Psi$ or the relative distribution
functional $\rho$ does not yield the probabilities of states by
themselves because they are not normalizable in the whole state space
$\H$. Some additional prescription must be given to make it physically
meaningful objects.  We utilize the dynamical foliation structure
of the Hilbert space given in the previous subspace for that purpose.

First let us define an \newterm{acausal subspace} as a linear manifold
$\L$ in $\H$ such that
\begin{itemize}
\item[i)] $\L$ is closed, i.e., $\bar\L=\L$,
\item[ii)] $\L$ is transversal to the null submanifold $\N$, i.e.,
$\L\cap\N=\{0\}$.
\end{itemize}
Then, as is clear from \Eq{WeakHamiltonianConstraint}, for any given
function $\Psi|_\L$ on an acausal subspace $\L$, we can find a
relative probability amplitude functional which satisfies the
weak hamiltonian constraints and coincides with $\Psi|_\L$ on
$\L$. Further it is unique on the submanifold $\L+\N$, which we
call the \newterm{development of $\L$} and denote it by $D(\L)$.
The situation is the same for the relative density functional $\rho$.
Thus acausal subspaces play the role of Cauchy surface for the weak
hamiltonian constraint.

This situation should be compared with that of the classical statistical
dynamics of a totally constrained system. There the relative distribution
function satisfies \Eqs{EqForRho1}{EqForRho2} and acausal submanifolds
which are transversal to the causal submanifolds play the role of Cauchy
surface\CITE{Kodama.H1995}. Thus there is a very good correspondence
between the classical theory and the quantum one. Hence, recalling that
each acausal submanifold corresponds to the set of states at some instant
and the restriction of the relative density distribution on that
acausal submanifold yields the probability distribution of states
at that instant in the classical theory, it is natural to require that
the relative probability amplitude $\Psi$ is bounded on acausal subspaces,
and to interpret its restriction on each acausal subspace as giving
the probability for each state contained in that acausal subspace after
normalization. This interpretation is also natural considering the fact
that $h_\alpha$ can be regarded as a generator of time translation which
moves any acausal subspace $\L$ transversally, i.e., $h_\alpha\L\cap\L=\{0\}$.
This implies that each acausal subspace corresponds to a set of states
at some instant.

This interpretation is mathematically formulated in the following way.
First for the relative probability amplitude $\Psi$ let us define
its norm on an acausal subspace $\L$ by
\Beq
||\Psi||_\L:=\sup_{u\in\L} |\Psi(u)|/||u||.
\Eeq
Then the probability for a state $u$ in $\L$ to be observed is given by
\Beq
\Pr(u;\L)=|\Psi(u)|^2/||\Psi||_\L^2.
\Eeq
Further for an operator $A$ such that $A\L\subseteq\L$ and its restriction
on $\L$ is a bounded self-adjoint operator, its expectation value on $\L$
is given by
\Beq
<A>_\L=\sum_n\overline{\Psi(u_n)}\Psi(Au_n)/||\Psi||_\L^2,
\label{Def:ExpectationValue:Pure}\Eeq
where $\{u_n\}$ is an orthonormal basis of $\L$. $<A>_\L$ does not depends
on the choice of the basis.

Note that since $\Psi$ is bounded on $\L$, there exists a unit vector
$\Phi_\L\in\L$ such that
\Beq
\Psi(u)/||\Psi||_\L=(\Phi_\L,u) \quad \forall u\in\L.
\Eeq
In terms of this vector the above equations are written exactly in the
same forms as in the ordinary quantum mechanics:
\Beqr
&&\Pr(u;\L)=|(\Phi_\L,u)|^2,\\
&&<A>_\L=(\Phi_\L,A\Phi_\L). \label{Def:ExpectationValue:Pure1}
\Eeqr

On the other hand when the dynamical state of the system is represented
by a relative density functional $R$, we require that its restriction
on $\L$ is of the trace-class functional and define its trace norm by
\Beq
||R||_\L:=\sum_n R(u_n,u_n).
\Eeq
Then the probability for $u\in\L$ is given by
\Beq
\Pr(u;\L)=R(u,u)/||R||_\L,
\Eeq
and for an operator $A$ of the same nature as above, its expectation
value by
\Beq
<A>_\L=\sum_n R(u_n,Au_n)/||R||_\L.
\Eeq
Further, due to the requirement on $R$, there exists an operator $\rho_\L$
with unit trace on $\L$ such that
\Beq
R(u,v)/||R||_\L=(u,\rho_\L v) \quad \forall u,v\in\L.
\Eeq
In terms of this operator $\Pr(u;\L)$ and $<A>_\L$ are written as
\Beqr
&&\Pr(u;\L)=(u,\rho_L u),\\
&&<A>_\L=\Tr_\L \rho_\L A,
\Eeqr
where $\Tr_\L$ denotes the trace in $\L$.
Thus each acausal subspace plays the role of the state space in the
ordinary quantum mechanics. In other words the whole state space in the
present formulation is a kind of web of the state spaces corresponding
to all the possible instants.

In practical situations each acausal subspace $\L$ is specified
as a common eigenspace of a set of projection operators $E_\alpha$
corresponding to time variables or instant functions in the classical
theory. Then
a quantity which can be simultaneously measurable with these operators
should correspond to an operator $A$ which commutes with $E_\alpha$. This
implies that $A\L\subseteq\L$.  This is the reason why we imposed
this condition on the operator in defining its expectation value on $\L$
above.

Mathematically each acausal subspace $\L$ is an eigenspace of the
single orthogonal projection operator
\Beq
E: \H \maps \L,
\Eeq
which will be called the \newterm{instant operator for $\L$}. This may
intuitively contradict the fact that classically we need
instant functions of the same number as that of the independent
constraint functions.  However, there exists no real contradiction because
any mutually commuting set of bounded self-adjoint operators can be
represented as functions of a single bounded self-adjoint operator
by von Neumann's theorem\CITE{Schwartz.J1967B}.

On the basis of the interpretation and normalization of $\Psi$ and
$R$ explained above the quantum dynamics of the totally constrained system
is formulated as follows. For simplicity we only consider the case
in which the dynamical state is represented by a relative probability
amplitude. The extension to the mixed-state cases is trivial.

First when a data set of measured values of physical quantities are
given, it determines a state $u_0$ in $\H$. Let $\L_0$ be an acausal
subspace such that $u_0\in\L_0$. Then the relative
probability amplitude $\Psi$ on $\L_0$ is uniquely determined modulo
a constant phase by the condition
\Beq
\Psi(u)=(u_0,u)/||u_0||^2 \quad u\in\L_0.
\label{InitialValueProblemForPsi}
\Eeq
By taking this functional as the initial value, the weak hamiltonian constraint
uniquely determines $\Psi$ on $D(\L_0)$.  Then for any instant operator
$E$ or the corresponding acausal subspace $\L$ such that $\L\in
D(\L_0)$ we can predict the probability for each state $u\in\L$ by
\Beq
\Pr(u;\L)=|\Psi(u)|^2/||\Psi||_{\L}^2,
\Eeq
and the expectation value of an operator $A$ such that $[A,E]=0$ by
\Beq
<A>_{\L}=\sum_n\overline{\Psi(u_n)}\Psi(Au_n)/||\Psi||_{\L}^2.
\Eeq

\subsection{Unitarity problem}

Though this formulation of dynamics is apparently well-posed, it
has various hidden difficulties.  First of all note that if $\N$ is closed,
the whole Hilbert space is orthogonally decomposed as the direct sum
$\H=\K\oplus\N$. Clearly $\K$ becomes a maximal acausal subspace.
Hence it is natural to require that $\Psi$ is bounded on $\K$. Then, however,
$\Psi$ becomes bounded on the whole Hilbert space, and it is represented
by some normalizable state vector $\Phi$. Further, since $\K$ coincides
with $\cap_\alpha\kernel{h_\alpha} $ from \Eq{NullSpace:def}, $\Psi$
satisfies the Dirac constraints $h_\alpha\Phi=0$($\forall \alpha$).
This implies that $\N$ should not be closed in order that $\Psi$(or
$R$) corresponds to the unnormalizable relative distribution function
in the classical theory.  This argument also shows that $\N$ should be dense
in $\H$, or equivalently $\K=\{0\}$ if we require that the weak hamiltonian
constraints have no solution that is bounded on the whole Hilbert space.

For example, let us consider an unconstrained canonical system with
a hamiltonian $h_0$ on the phase space $\RF^{2n}(\ni(\bm{x},\bm{p}))$
and its embedding into a totally constrained system with a constraint
$h=p_0+h_0$ on the phase space $\RF^{2n+2}(\ni(x^0,\bm{x},p_0,\bm{p}))$.
Then in the coordinate representation, the state space $\H_0$ of the
quantized unconstrained system is given by $\H_0=L_2(\RF^n)$ and
that of the quantized constrained system by a direct integral
\Beq
\H=L_2(\RF^{n+1})\cong \int_\sRF \oplus \H_0(x^0) dx^0,
\Eeq
where $\H_0(x^0)=\H_0$. In this representation $\Phi\in\H$ is expressed
as
\Beq
\Phi=\int_\sRF\oplus \Phi(x^0) dx^0,
\Eeq
and the Dirac constraint by the Schr\"{o}dinger equation
\Beq
i\partial_t \Phi(t)=h_0 \Phi(t).
\Eeq
Hence if $\Phi$ satisfies the Dirac hamiltonian constraint, its norm diverges:
\Beq
(\Psi,\Psi)=\int_\RF (\Phi(x^0),\Phi(x^0))dx^0
=(\Phi(0),\Phi(0))\int_\RF dx^0=+\infty.
\Eeq
This implies that $\kernel{h}=\{0\}$.

The unclosed nature of $\N$ has profound implications and is the most
cumbersome aspect of the present formulation. Firstly, though the solutions
to the weak hamiltonian constraint are one-to-one correspondence with
the functions on the quotient space $\H/\N$, this reduction does not
give a useful information because $\H/\N$ is not even Hausdorff with
respect to the natural quotient topology. This is the reason why we
used acausal subspaces to normalize $\Psi$(and $R$).
Secondly it implies that there does not exist a maximal acausal subspace
in general.  Though we cannot show it generally, we can prove it exactly
at least for the totally constrained system with a single constraint.

\begin{proposition}
If $h$ is a closed operator such that $\kernel{h}=\{0\}$ and $\N=
\range{h}$ is not closed, there does not exist a closed subspace
$\M$ such that $\H=\M+\N$ and $\M\cap\N=\{0\}$.
\end{proposition}
\begin{proof}
Suppose that there existed a closed subspace $\M$ with the properties
in the proposition and let us consider the natural projection
$P_\orth:\H\maps\M_\orth$ with respect to the orthogonal decomposition
of $\H$ given by
$$
\H=\M\oplus \M_\orth.
$$
Then any $u\in\M_\orth$ is  written as $u=v+w$ with
$v\in\M$ and $w\in\N$, hence $w=u-v$. This implies that $u=P_\orth w$.
Hence $P_\orth(\N)=\M_\orth$. Further if $P_\orth u=0$ for $u\in\N$,
then $u\in\M$. However, since $\M\cap\N=\{0\}$, this implies that $u=0$.
Hence the restriction of $P_\orth$ on $\N$ yields a continuous
one-to-one mapping on to $\M_\orth$.

On the other hand, since $h$ is a closed operator, its graph in $\H\oplus\H$
is given by a closed subspace $\tilde \N$, and for the natural projection
$P_2$ on the second component $P_2(\tilde\N)$ coincides with $\N$. Further
since $\kernel{h}=\{0\}$, the restriction of $P_2$ on $\tilde N$ is
one-to-one. Hence the restriction of the mapping $P_\orth P_2$ on
$\tilde N$ yields a continuous bijection onto $\M_\orth$. However,
since $\M_\orth$ is closed, this bijection has a continuous inverse $f$
from the Inverse Mapping Theorem\CITE{Conway.J1985B}.
Clearly $P_2 f$ is a continuous
bijection from $\M_\orth$ onto $\N$, and yields the inverse of $P_\orth|_\N$.
This implies that $\N$ and $\M_\orth$ is isomorphic, hence $\N$
is closed. This contradicts the assumption. \QED
\end{proof}

\bigskip

\begin{corollary}
If $h$ is a self-adjoint operator such that $\range{h}$ is not closed,
the same conclusion as in the theorem holds even if $\kernel{h}\not=\{0\}$.
\end{corollary}
\begin{proof}
If $\kernel{h}\not=\{0\}$, under the orthogonal decomposition
$$
\H=\kernel{h}\oplus\H_\orth,
$$
$\domain{h}$ is decomposed as
$$
\domain{h}=\kernel{h}\oplus \D.
$$
Suppose that $\H$ were decomposed in terms of a closed subspace $\M$ as a
direct sum
$$
\H=\M+\N.
$$
Then, since $\N\subset\H_\orth$ and $\overline{\range{h}}=(\kernel{h})_\orth$,
the orthogonal projection
$P:\H\maps\kernel{h}$ maps $\M$ onto $\kernel{h}$. Hence $\M$ is
orthogonally decomposed as
$$
\M=\M_0\oplus (\M\cap\H_\orth),
$$
where the restriction of $P$ on $\M_0$ is one-to-one onto $\kernel{h}$,
and any vector $u$ in $\H_\orth$ is written as a direct sum
$$
u=u_0+u_1+u_2,
$$
where $u_0\in\M_0$, $u_1\in\M\cap\H_\orth$ and $u_2\in\N$. However,
since $P|_{\M_0}$ is injective and $P(u_0)=P(u)=0$, it follows that
$u_0=0$. Hence $\H_\orth$ is written as a direct sum
$$
\H_\orth=\M\cap\H_\orth + \N.
$$
Obviously $\M\cap\H_\orth$ is closed.
However, since $\overline{\range{h}}=(\kernel{h})_\orth$, the restriction
of $h$ on $\H_\orth$ yields a self-adjoint operator on $\H_\orth$,
while such a closed subspace does not exists by the proposition. \QED
\end{proof}

The non-existence of a maximal acausal subspace implies that for any choice
of the initial acausal subspace $\L_0$ we can determine the relative
probability amplitude only on a proper subset of $\H$ from a normalized
initial data on $\L_0$.  Of course, since we can find a sequence of
acausal subspaces whose developments contain the whole $\H$ in the limit,
this will not be a serious problem if the prediction does not depend
on the choice of the initial acausal subspace as far as it contains
the vector corresponding to the given data set. Unfortunately, however,
it is not the case.  On the contrary the prediction depends on the
choice of the initial acausal subspace.  For example, let $u_0$ be
an acausal vector given by the initial data set, and pick up a vector $n$
in $\N$ such that $(u_0,n)\not=0$. If $u_0$ is chosen so that it is
orthogonal to $\cap_\alpha\kernel{h_\alpha}$, such a vector $n$ always
exists. Further let $u_1$ be a vector such that the two-dimensional
subspace $\L_1$ spanned by $u_0$ and $u_1$ is acausal. Then the
subspace $\L_2$ spanned by $u_0$ and $u_2=u_1+n$ is also acausal
and shares the vector $u_0$ with $\L_1$. Let the solution
to the weak hamiltonian constraints determined by the condition
\Eq{InitialValueProblemForPsi} with $\L_0$ replaced by $\L_1$ be $\Psi_1$,
and the corresponding solution for $\L_2$ be $\Psi_2$. Then
\Beq
\Psi_2(u_2)=(u_0,u_2)/||u_0||^2=(u_0,n)/||u_0||^2\not=0,
\Eeq
while $\Psi_1(u_2)=\Psi_1(u_1+n)=\Psi_1(u_1)=0$.  Thus the predictions
are inconsistent.

This inconsistency is closely related with breakdown of unitarity.
To see this, let us consider two acausal subspaces $\L_1$ and $\L_2$.
For each vector $u$ in $\L_1\cap D(\L_2)$ there exists a unique vector
$\Theta(u)$ in $\L_2\cap D(\L_1)$ such that $\Theta(u)-u\in\N$ from
the condition ii) on acausal subspaces.  Hence we obtain a bijective
linear mapping
\Beq
\Theta: \Mapping{\L_1\cap D(\L_2)}{u}{\L_2\cap D(\L_1)}{\Theta(u)}.
\Eeq
Let us call this mapping a \newterm{causal mapping}. From this definition
for any relative probability functional $\Psi$ which satisfies the weak
hamiltonian constraint and defined on $D(L_1)\cap D(L_2)$ the following
equality holds:
\Beq
\Psi(\Theta(u))=\Psi(u) \quad u\in\L_1\cap D(\L_2).
\Eeq

In general $\Theta(u)$ is not continuous. Hence even if $\Psi$ is
bounded on $\L_2\cap D(\L_1)$, it may not be bounded on $\L_1\cap D(\L_2)$.
Clearly the probability interpretation break down in such situations.

For example, let us consider the quantum totally constrained system
with the Hilbert space $\H=L_2(\RF^2)\ni(t,x)$ and the constraint
operator $h=-i\partial_t$. Let $\phi_n(x)$ be the orthonormal basis
of $L_2(\RF)$ and let $\chi_n(t)$ be the series of functions of the form
\Beq
\chi_n(t)={c_n\over |t|^{1+{1\over n^2}}+1},
\Eeq
where $c_n$ is the positive constant determined by the condition
\Beq
\int_\RF dt \chi_n(t)^2=1.
\Eeq
Then $\lim_{n\tend\infty}c_n=1$, and
\Beq
\lambda_n:=\int_\RF dt \chi_n(t)> n^2c_n.
\Eeq
In terms of these functions let us define the subspace $\L_1$ of $\H$ by
\Beq
\L_1:=\SetDef{u=\sum_n a_n \chi_n(t)\phi_n(x)\in L_2(\RF^2)}
{a_n\in\CF}.
\Eeq
Then it is easily checked that this subspace is acausal and
isomorphic to the Hilbert space $l_2=\{\bm{a}=(a_1,a_2,\cdots)\}$ with
norm $||\bm{a}||^2=\sum_n|a_n|^2$. Further in terms of a function $\chi(t)$
such that
\Beq
\lambda:=\int_\RF dt\chi(t) \not=0, \quad
\int_\RF dt\chi(t)^2<+\infty,
\Eeq
define the subspace $\L_2$ by
\Beq
\L_2:=\SetDef{u=\chi(t) \sum_n b_n \phi_n(x)\in L_2(\RF^2)}
{b_n\in\CF}.
\Eeq
Then this is also an acausal subspace and isomorphic to the Hilbert
space $l_2=\{\bm{b}=(b_1,b_2,\cdots)\}$. The causal mapping from $\L_1$
to $\L_2$ is expressed in terms of $\bm{a}$ and $\bm{b}$ as
\Beq
\bm{b}=\Theta(\bm{a}) \equivalent
b_n={\lambda_n\over\lambda}a_n.
\Eeq
Hence the domain of $\Theta$ is given by
\Beq
\L_1\cap D(\L_2)=\SetDef{\bm{a}\in l_2}{\sum_n\lambda_n^2 |a_n|^2<+\infty},
\Eeq
and the range of $\Theta$ coincides with $\L_2$. It is easy to see that
the linear mapping $\Theta$ is not bounded. Further the linear functional
$\Psi_2$ on $\L_2$ defined by
\Beq
\Psi_2(\bm{b})=\sum_n n^{-1}b_n
\Eeq
is bounded and its norm is given by $\pi/\sqrt{6}$, but the function $\Psi_1$
on $\L_1\cap D(\L_2)$ defined as the pullback of $\Psi_2$ by $\Theta$ is
not bounded.  In fact for the sequence of unit vectors $\bm{a}_m$ in
$\L_1\cap D(\L_2)$ given by $(\bm{a}_m)_n:=\delta_{m,n}$, its value diverges
as
\Beq
\Psi_1(\bm{a}_m)=\Psi_2(\Theta(\bm{a}_m))={\lambda_m\over m\lambda}
\tend \infty (m\tend\infty).
\Eeq

If we consider only the acausal subspaces for which the causal mappings
among them are continuous with respect to the weak topology, this
disastrous violation of unitarity is avoided. However, even under
such restriction, there may still occur a physically uncomfortable
phenomenon. To see this, recall the simple example of the two-dimensional
acausal subspaces discussed above. In this example the causal mapping
is given by
\Beq
\Theta(u_0)=u_0,\quad \Theta(u_1)=u_2,
\Eeq
and continuous because it is finite dimensional. Let us normalize $u_0$
and $u_1$ to unit vectors, and consider a one-parameter family of linear
functionals with unit norm on $\L_1$ defined by
\Beq
\Psi_\phi(u)=(u_0\cos\phi+u_1\sin\phi,u).
\Eeq
Then the norms of their extensions on $\L_2$ are given by
\Beq
||\Psi_\phi||^2_{\L_2}={||u_2||^2\cos^2\phi+\sin^2\phi
-(u_2,u_0)\sin\phi\cos\phi
\over ||u_2||^2-|(u_0,u_2)|^2}.
\Eeq
Hence they do not coincide with each other unless $||u_2||=1$ and
$(u_0,u_2)=0$, in other words, unless $\Theta$ is a unitary mapping.
As expected from this example, even if the norm of $\Psi$ satisfying
the weak hamiltonian constraint is normalized to be
unity on one acausal subspace $\L_1$, its norm on another acausal
subspace $\L_2$ may change depending on its initial value on $\L_1$.
Though this phenomenon may not be regarded as serious for pure
dynamical states, it will cause a trouble when we consider mixed
dynamical states. This pathology is avoided only when the causal
mapping is conformal, i.e., $||\Theta(u)||/||u||$ is constant
as shown in the following proposition.

\begin{proposition}
For a bijective linear mapping between two Hilbert spaces
$\Theta:\H_1\maps\H_2$ such that $\Theta$ and
$\Theta^{-1}$ are both weakly continuous, its dual mapping $\Theta^*:
\H_2^*\maps\H_1^*$ is conformal if and only if $\Theta$ is conformal.
\end{proposition}
\begin{proof}
$\Theta^*$ is obviously conformal if $\Theta$ is conformal.
To show the inverse, suppose that $||\Theta^*\Psi||=c||\Psi||$
for some positive constant $c$. Let $\Phi$ be the vector such that
$\Psi(u)=(\Phi,u)$. Then from the definition of the norm it follows
that
$$
{|\Theta^*\Psi(x)|\over||x||}=|(\Phi,{\Theta(x)\over||x||})|
\leq c||\Psi||.
$$
In particular for the functional corresponding to
$\Phi={\Theta(x)\over||x||}$, we obtain
$$
{||\Theta(x)||\over||x||}\le c \quad \forall x\in\H_1.
$$
By the same argument for $\Theta^{-1}$ we obtain
$$
{||x||\over||\Theta(x)||}\le {1\over c} \quad \forall x\in\H_1.
$$
These two equations are consistent only when
$$
{||\Theta(x)||\over||x||}= c \quad \forall x\in\H_1.
$$
This proves the proposition.  \QED
\end{proof}

These observations show that we must find some prescription to
pick up special families of acausal subspaces such that causal
mapping among them are always conformal in order for the formalism
to give a physically sensible dynamics.  In the next section
we will show that there actually exists such a prescription.

\section{Physical Acausal Subspaces}

In the ordinary quantum mechanics the algebra of all the bounded
operators coincides with the algebra generated by all the operators
corresponding to constants of motion. Hence, in order for an acausal
subspace $\L$ to play the role of a physical state space at some
instant, it should be large enough so that all the operators
corresponding to constants of motion can be measurable on $\L$ and
they separate vectors in $\L$. This is equivalent to require that the
algebra $\C$ of all the constants of motion leaves $\L$ invariant and
its representation on $\L$ is irreducible. Though this appears to be a
natural prescription to pick up physical acausal subspaces, it does
not work in the present formalism because $\C$ generally contains some
of the constraint operators, which do not keep acausal subspaces
invariant.

This defect of the prescription can be eliminated by replacing $\C$ by
$\B$ in the prescription if we can decompose $\C$ into a subalgebra
$\Z$ generated by the constraint operators contained in $\C$ and
another subalgebra $\B$ which corresponds to the set of quantities
obtained by restricting the constants of motion on the constraint
submanifold in the classical theory. This modification of the
prescription is natural because $\B$ corresponds to the algebra of
functions on the reduced phase space obtained by solving the
hamiltonian constraints in the classical theory. In this section we
will show that the central decomposition of the von Neumann algebra
(or $\VNA$-algebra) $\C$ of constants of motion naturally leads to such a
decomposition, and if $\C$ is type I, our formulation gives a consistent
dynamics with respect to the physical acausal subspaces selected by the
prescription above under some additional assumptions.

\subsection{The central decomposition of $\VNA$-algebra $\C$ and physical
subalgebra}

We define the algebra $\C$ of the constants of motion as the set of
bounded operators which commute with all $h_\alpha$,
\Beq
\C:=\{h_\alpha\}',
\label{C:def}\Eeq
and the algebra of constraint operators $\W$ by
\Beq
\W:=\C'.
\Eeq
Then they become $\VNA$-algebras and $\C$ coincides with the commutant
of $\W$ by the double-commutant theorem\CITE{Conway.J1985B}:
\Beq
\C=\W'.
\Eeq
Note that we are considering not the weak constants of motion but the
exact ones. Hence $\W$ is not contained in $\C$ unless all the
constraint operators commute with each other.

When $h_\alpha$ are not bounded, definition \Eq{C:def} may be unclear.
For such cases, in terms of the spectral decomposition of $h_\alpha$,
\Beq
h_\alpha=\int_\sRF \lambda dE_\alpha(\lambda),
\label{SpectralDecompositionOfh}\Eeq
we define $\C$ by
\Beqr
&&\C:=S';\\
&&S:=\bigcup_\alpha \SetDef{E_\alpha(s)}{s:\r{Borel\ set\ of}\;\RF}.
\Eeqr
Then $\W$ coincides with the $\VNA$-algebra generated by $S$:
\Beq
\W=\VNA(S).
\Eeq

If we define a set of bounded self-adjoint operators $H_\alpha$
in terms of a bounded monotonic function $f$ on $\RF$ and $h_\alpha$
by
\Beq
H_\alpha=f(h_\alpha):=\int_\sRF f(\lambda) dE_\alpha(\lambda),
\label{BoundedConstraintOperators}\Eeq
$H_\alpha$ has the same spectral measure as the corresponding
$h_\alpha$.  Hence we can also express $\C$ and $\W$ as
\Beqr
&&\C=\{H_\alpha\}',\\
&&\W=\VNA(\{H_\alpha\}).
\Eeqr

The set of the common elements of $\C$ and $\W$ coincides with the
center of $\C$(and $\W$):
\Beq
\Z=\C\cap\W.
\Eeq
As is well-know, the center $\Z$ is generated by a single self-adjoint
operator and in terms of its spectral decomposition the algebra $\C$
and $\W$ are written as direct integrals\CITE{Schwartz.J1967B}. Let us
denote the generating operator of $\Z$ by $\Lambda$,
\Beq
\Z=\VNA(\Lambda),
\Eeq
and the scalar-valued spectral measure with support on the spectrum
$\sigma(\Lambda)(\subset\RF)$ by $\mu$. Then the Hilbert space is
represented by a direct integral of Hilbert spaces
$\H(\lambda)(\subseteq\hat\H_\infty)$ as
\Beq
\H \cong \int_{\sigma(\Lambda)}\oplus\H(\lambda)d\mu(\lambda),
\label{CentralDecomposition1}\Eeq
and $\C$ and $\W$ are decomposed into direct integrals as
\Beqr
&&\C \cong \int_{\sigma(\Lambda)}\oplus\C(\lambda)d\mu(\lambda),
\label{CentralDecomposition2}\\
&&\W \cong \int_{\sigma(\Lambda)}\oplus\W(\lambda)d\mu(\lambda),
\label{CentralDecomposition3}
\Eeqr
where $\C(\lambda)$ and $\W(\lambda)$ are $\VNA$-algebra on
$\H(\lambda)$, and related by $\W(\lambda)=\C(\lambda)'$.

The most important property of this central decomposition is the fact
that $\C(\lambda)$ and $\W(\lambda)$ become factors, i.e., their
centers become trivial. Let us denote the spatial isomorphism classes
which appear in $\{\C(\lambda)\}$ by $\SetDef{\hat \B_j}{j\in J}$, and
define the subsets $\sigma_j$ of $\sigma(\Lambda)$ by
\Beq
\sigma_j:=\SetDef{\lambda\in\sigma(\Lambda)}{\C(\lambda)\cong\hat \B_j}.
\Eeq
Then $\H$, $\C$ and $\W$ are decomposed into direct sums as
\Beqr
&&\H\cong \oplus_{j\in J}L_2(\sigma_j;\mu)\bar\otimes \hat \H_j,
\label{FactorDecomposition1}\\
&&\C\cong \oplus_{j\in J}L_\infty(\sigma_j;\mu)\bar\otimes \hat \B_j,
\label{FactorDecomposition2}\\
&&\W\cong \oplus_{j\in J}L_\infty(\sigma_j;\mu)\bar\otimes \hat \W_j,
\label{FactorDecomposition3}\\
&&\Z\cong L_\infty(\sigma(\Lambda);\mu)\cong
\oplus_{j\in J}L_\infty(\sigma_j;\mu)\otimes 1,
\label{FactorDecomposition4}
\Eeqr
where $L_2(\sigma_j;\mu)$ and $L_\infty(\sigma_j;\mu)$ are the sets of
$L_2$-functions and bounded functions on $\sigma_j$ with respect to
the measure $\mu|_{\sigma_j}$, respectively, and the relations
\Beq
\W_j=\B_j', \quad  \B_j\cap \W_j=\CF
\Eeq
hold.

Therefore if we define $\B$ by
\Beq
\B\cong \oplus_{j\in J}(1\otimes\hat\B_j),
\label{GeneralPhysicalSubalgebra}\Eeq
$\C$ is written as
\Beq
\C\cong \Z\bar\otimes\B.
\Eeq
Thus the central decomposition naturally yields a subalgebra of $\C$
which corresponds to the set of constants of motion restricted on the
constraint submanifold in the classical theory.

On the basis of this fact and the argument at the beginning of this
section, we call the subalgebra $\B$ a \newterm{physical subalgebra}
of $\C$, and an acausal subspace $\L$ such that $\B\L=\L$
a \newterm{physical acausal subspace}.

Here note that the physical subalgebra is not unique due to the
spatial-isomorphism freedom in the central decomposition
representation.  This freedom is expressed as $U\B U^{-1}$ in terms of
a decomposable unitary operator $U$,
\Beq
U \cong \int_{\sigma(\Lambda)}\oplus U(\lambda)d\mu(\lambda).
\Eeq
This decomposability condition is satisfied if and only if $U\in
\Z'$. In the present case, however, this condition is not sufficient
and the additional condition $U\B U^{-1}\subseteq\C$ must be satisfied.
Here note that $\Z'=\VNA(\C,\W)$ from the following proposition.

\begin{proposition}
For a $\VNA$-algebra $\A$, the commutant of the center $\Z$ of $\A$
is generated by $\A$ and $\A'$:
$$
\Z'=\VNA(\A,\A').
$$
\end{proposition}
\begin{proof}
Clearly $\Z'\subseteq\VNA(\A,\A')$ and $\VNA(\A,\A')'\subseteq\Z$.
{}From the latter and the double commutant theorem we obtain
$$
\Z'\subseteq\VNA(\A,\A')''=\VNA(\A,\A').
$$
Hence $\Z'=\VNA(\A,\A')$.
\QED
\end{proof}

\subsection{Linear null submanifold}

In order to investigate the properties of acausal physical subspaces,
we need to know the relation between the linear null submanifold and
the $\VNA$-algebras defined in the previous subsection. For that
purpose let us introduce the operator algebra $\I$ defined by
\Beq
\I:=\SetDef{\sum_\alpha h_\alpha A_\alpha(\r{finite\ sum})}
{A_\alpha, h_\alpha A_\alpha \in\W},
\Eeq
and redefine the linear null submanifold $\N$ by
\Beq
\N:=\I \H.
\Eeq

In order to show that this null submanifold coincides with that
defined in the previous section, we first show that we can replace
$h_\alpha$ in these definitions by the bounded self-adjoint operators
$H_\alpha$ introduced above if we take $f$ properly.

\begin{proposition}\label{prop:BoundedConstraintOperators}
There exists self-adjoint generators $H_\alpha$ in $\W$ which satisfy
the following conditions:
\begin{itemize}
\item[i)] $\kernel{H_\alpha}=\kernel{h_\alpha}$,
\item[ii)] for any set of bounded Borel functions $f_\alpha$ such that
$f_\alpha(\lambda)$ is smooth in a neighborhood of $\lambda=0$,
$f_\alpha(0)=0$, and $f'_\alpha(0)\not=0$, $\I$ is written as
$$
\I=\sum_\alpha f_\alpha(H_\alpha)\W \;\r{(finite\ sum)}.
$$
\end{itemize}
\end{proposition}
\begin{proof}
In definition (\ref{BoundedConstraintOperators}) of $H_\alpha$
let us put
$$
f(\lambda)=\tanh(\lambda).
$$
Further let us define the self-adjoint operator $G_\alpha$ defined by
$$
G_\alpha:=\int_\sRF {\lambda\over f(\lambda)}dE_\alpha(\lambda).
$$
Then its inverse $G_\alpha^{-1}$ belongs to $\W$, and for the unitary
operator $U_\alpha$ defined by
$$
U_\alpha:= E_\alpha(\lambda\ge0)-E_\alpha(\lambda<0),
$$
the difference of $U_\alpha h_\alpha$ and $G_\alpha$,
$$
G_\alpha-U_\alpha h_\alpha=\int_\sRF \lambda{1-|\tanh\lambda|
\over \tanh\lambda}dE_\alpha(\lambda),
$$
is a bounded operator in $\W$.  In particular it follows from this that
$\domain{G_\alpha}=\domain{h_\alpha}$. Further it satisfies the relation
$$
h_\alpha=H_\alpha G_\alpha.
$$
Hence if $h_\alpha A_\alpha\in\W$, then $G_\alpha A_\alpha\in\W$ and
$h_\alpha A_\alpha=H_\alpha G_\alpha A_\alpha$. This implies $\I\subseteq
\sum H_\alpha \W$(finite sum). Conversely if $A_\alpha\in\W$, then $H_\alpha
A_\alpha=h_\alpha G_\alpha^{-1}A_\alpha\in\I$. Hence
$\sum_\alpha H_\alpha\W$(finite sum)$\subseteq\I$. Therefore
$H_\alpha$ satisfies i) and ii) for the case $f_\alpha(\lambda)=\lambda$.

Next let $f_\alpha$ be functions satisfying the conditions in ii).
Then, since $\spectrum{H_\alpha}$ is bounded,
$f_\alpha(H_\alpha)H_\alpha^{-1}$ is a bounded self-adjoint operator
and has a bounded inverse. From this it immediately follows that
ii) holds for $f_\alpha$.  \QED
\end{proof}

With help of this proposition we can easily show that $\N$ coincides
with the original one as follows.

\begin{corollary}
$\N=\sum_\alpha h_\alpha$(finite sum).
\end{corollary}
\begin{proof}
{}From the definition of $\I$ it immediately follows that $\N=\I\H\subseteq
\sum_\alpha \range{h_\alpha}$ (finite sum). On the other hand for the operators
$H_\alpha$ and $G_\alpha$ in the proof of the proposition and for $u\in
\range{h_\alpha}$ there exists a vector $v\in\domain{h_\alpha}$ such
that $u=h_\alpha v=H_\alpha G_\alpha v$. Hence if we put $w=G_\alpha
v$, we get $u=H_\alpha w$. Thus $u\in \I\H$, which implies that
$\sum_\alpha \range{h_\alpha}$(finite sum)$\subseteq \I\H=\N$.
\QED
\end{proof}

\subsection{Direct-sum decomposition of acausal subspaces}

In general a physical subalgebras is written as a direct sum of
factors which are not mutually spatially isomorphic. We show that
such decomposition induces a direct sum decomposition of physical
acausal subspaces.

We first prove the following basic proposition.

\begin{proposition}\label{prop:DecompositionOfAcausalSubspace}
For an orthogonal decomposition $\H=\H_1\oplus\H_2$,
\begin{itemize}
\item[1)]  $\C$ is decomposable as $\C=\C_1\oplus\C_2$ if and only if $\W$ is
decomposable as $\W=\W_1\oplus\W_2$.
\item[2)] Under this decomposition of the algebras the linear null
submanifold is decomposed as $\N=\N_1\oplus\N_2$. Further for
closed subspaces $\L_j\subseteq\H_j$($j=1,2$), $\L=\L_1\oplus\L_2$
is acausal if and only if $\L_j\cap\N_j=\{0\}$.
\end{itemize}
\end{proposition}
\begin{proof}
\noindent
1) Let $E:\H\maps\H_1$ be the orthogonal projection operator. Then
if $\C=\C_1\oplus\C_2$, $E\in\C=\W'$. On the other hand,
since $\C\H_j\subseteq\H_j$($j=1,2$), $E\in\C'=\W$. Hence
$\W_1:=E\W E=\W E$ and $\W_2:=(1-E)\W (1-E)=\W(1-E)$ are both
$\VNA$-subalgebra of $\W$, and for any $A\in\W$,
$$
A=AE+A(1-E)=EAE\oplus(1-E)A(1-E).
$$
Hence $\W$ is decomposed as $\W=\W_1\oplus\W_2$. The converse
is proved in the same way.

\noindent
2) If $\W=\W_1\oplus\W_2$, then $\W\H_j\subseteq\H_j$($j=1,2$),
hence $\I\H_j\subseteq\H_j$. From this it immediately follows that
$$
\N=\I\H=\I\H_1\oplus\I\H_2\equiv\N_1\oplus\N_2.
$$
The latter half is obvious since $\L\cap\N=(\L_1\cap\N_1)\oplus(\L_2\cap\N_2)$.
\QED
\end{proof}

For acausal subspaces invariant under a decomposable subalgebra of $\C$,
the following stronger proposition holds.

\begin{proposition}\label{prop:DecompositionOfPhysicalAcausalSubspace}
Assume that $\C$ is decomposable as $\C=\C_1\oplus\C_2$ under the
orthogonal decomposition $\H=\H_1\oplus\H_2$, and let $\A$ be
a subalgebra which is decomposable as $\A=\A_1\oplus\A_2$.
Then if $\L$ is an acausal subspace such that $\A\L\subseteq\L$,
$\L_j=\L\cap\H_j$($j=1,2$) is an acausal subspace invariant under
$\A_j$ and $\L$ is written as $\L=\L_1\oplus\L_2$. Conversely
if $\L_j$($j=1,2$) is an acausal subspace of $\H_j$ such that
$\A_j\L_j\subseteq\L_j$, $\L=\L_1\oplus\L_2$ is an acausal
subspace invariant under $\A$.
\end{proposition}
\begin{proof}
Let $\L$ be an acausal subspace invariant under $\A$. Then
$$
\L=(\A_1\oplus\A_2)\L=\A_1\L\oplus\A_2\L.
$$
Obviously $\L_j:=\A_j\L=\L\cap\H_j$ is a closed subspace invariant
under $\A_j$ and is acausal, Conversely if $\L_j$ is an acausal subspace
of $\H_j$ invariant under $\A_j$, $\L=\L_1\oplus\L_2$ is acausal
 from the previous proposition and invariant under $\A$.
\QED
\end{proof}

This proposition shows that the study of physical acausal subspaces
is reduced to that for each factor $\hat\B_j$ in $\H_j$ in the
decomposition \Eqs{FactorDecomposition1}{GeneralPhysicalSubalgebra}.
Hence from now on we only consider the case in which $\B$ is a factor.

Further from this proposition we obtain the following corollary.

\begin{corollary}\label{cor:SeparationOfNullSector}
Let $\H$, $\C$ and $\A$ be the same as in the proposition.  If
$\H_2\subseteq\N$, an acausal subspace $\L$ invariant under $\A$
is contained in $\H_1$.
\end{corollary}

Under the spectral decomposition \Eq{CentralDecomposition3} each
bounded constraint operator $H_\alpha$ is represented by a direct
integral as
\Beq
H_\alpha=\int_{\sigma(\Lambda)}\oplus H_\alpha(\lambda)d\mu(\lambda).
\Eeq
In particular if $H_\alpha$ is contained in $\Z$, $H_\alpha(\lambda)$
reduces to a function. Let $\chi(\lambda)$ be the characteristic function
of an open neighborhood $\U$ of the zero
points of the function $H_\alpha(\lambda)$, and $E$ be the orthogonal
projection operator defined by
\Beq
E:=\int_{\sigma(\Lambda)}\oplus \chi(\lambda)d\mu(\lambda).
\Eeq
Then $\H$ is decomposed as $\H=E\H\oplus(1-E)\H$ and $H_\alpha$ is
invertible on $(1-E)\H$. Hence $(1-E)\H\subseteq\N$. Therefore
if there exists $H_\alpha$ contained in $\Z$, the above corollary
implies that the physical acausal subspaces are determined by
the structure of $\C(\lambda)$ in a neighborhood of the common
zero points of such $H_\alpha(\lambda)$'s. Of course this simplification
does not occur in the case  $Z$ does not contain $H_\alpha$ even if $Z$
is non-trivial.

With the help of this corollary we can show that quantum dynamics
becomes trivial if the linear null submanifold is closed.
To see this, let $\K$ be the closed subspace of $\H$ defined by
\Beq
\K:=\bigcap_\alpha \kernel{h_\alpha},
\Eeq
Then as discussed in the previous section, vectors in $\K$ satisfy
the standard Dirac constraints and dynamics is frozen on $\K$.
To be precise, the following proposition holds.

\begin{proposition}
Let $E:\H\maps\L$ be the instant operator for an acausal subspace $\L$.
Then for any bounded constraint operator $H_\alpha$ in
\Prop{prop:BoundedConstraintOperators} the following equivalence
holds:
$$
EH_\alpha=H_\alpha E \equivalent \L\subseteq\kernel{H_\alpha}.
$$
In particular if $\L\not\subseteq\K$, there exists $H_\alpha$
such that $[E,H_\alpha]\not=0$.
\end{proposition}
\begin{proof}
If $EH_\alpha=H_\alpha E$, then
$$
EH_\alpha\L=H_\alpha\L\subseteq \N\cap\L=\{0\},
$$
hence $\L\subseteq\kernel{H_\alpha}$. Conversely if
$\L\subseteq\kernel{H_\alpha}$, $H_\alpha E=0$. Hence from the
self-adjointness of $H_\alpha$ and $E$, we obtain $H_\alpha E
=EH_\alpha$. \QED
\end{proof}

Since $\W$ becomes trivial on $\K$, under the orthogonal decomposition
\Beq
\H=\K\oplus\H_0,
\Eeq
$\W$ is decomposed as
\Beq
\W=\CF\oplus\W_0.
\Eeq
Hence from \Prop{prop:DecompositionOfAcausalSubspace} $\C$ is
decomposed as
\Beq
\C=\B(\K)\oplus\C_0.
\Eeq
Therefore from \Prop{prop:DecompositionOfPhysicalAcausalSubspace} any
physical acausal subspace $\L$ such that $\L\not\subseteq\H_0$ is
written as
\Beq
\L=\K\oplus\L_0.
\Eeq

On the other hand the closure of $\N$ coincides with the orthogonal
complement $\K_\orth$ from the following proposition.

\begin{proposition}
Let $E_\K:\H\maps\K$ be the orthogonal projection operator. Then
$$
\bar\I=(1-E_\K)\W.
$$
In particular
$$
\bar\N=\K_\orth.
$$
\end{proposition}
\begin{proof}
Let $A$ be a bounded self-adjoint operator, and its spectral measure
be $E(\alpha)$. Then for any $x\in\kernel{A}^\orth$ and any positive
number $\epsilon$, there exists a positive number $\delta$ such that
$$
\int_{|\alpha|<\delta}(x,dE(\alpha)x)<\epsilon.
$$
Let $P_n(\alpha)$ be a sequence of polynomials such that
$0\le P_n(\alpha)\le2$, $P_n(0)=0$, and $P_n(\alpha)$ converges
to unity locally uniformly on $\spectrum{A}-{0}$. Then
for the same $x$ as above we obtain
\begin{eqnarray*}
&||(1-P_n(A))x||^2&=\int_\sRF (1-P_n(A))^2(x,dE(\alpha)x)\\
&&\le \epsilon+\int_{|\alpha|\ge\delta} (1-P_n(A))^2(x,dE(\alpha)x)\\
&&\le \epsilon+\max_{|\alpha|\ge\delta}(1-P_n(A))^2||x||^2.
\end{eqnarray*}
The right-hand side of this equation tends to $\epsilon$ in the
$n\tend\infty$ limit. However, since $\epsilon$ is arbitrary, this
means that the left-hand side converges to zero. Hence we obtain
$$
\SOT-\lim_{n\tend\infty} P_n(A)=1-E_A,
$$
where $E_A$ is the orthogonal projection operator onto $\kernel{A}$.

Now let us apply this result to operators in $\W$.  First note that
there exists a countable subset of bounded constraint operators
$\{H_j\}$($j=1,\cdots,m\le+\infty$) such that
$\K=\cap_{j=1}^m\kernel{H_j}$ since we are assuming that $\H$ is
separable.  Let $Q_l$ be ${\sum_{j=1}^l H_j^2}$($l\le m,l<\infty$),
and $E_l$ be the orthogonal projection operator onto
$\K_l=\kernel{Q_l}$. Then since $P_n(Q_l)\in\I$, $1-E_l\in\bar\I$ from
the result above. However, since $\K_l=\cap_{j=1}^l\kernel{H_j}$ and
$E_{l+1}\le E_l$, by taking the limit $l\tend+\infty$ when
$m=+\infty$, we obtain $1-E_\K=1-E_m\in\bar\I$.  Hence
$(1-E_\K)\W\subseteq\bar\I$. On the other hand clearly $E_\K\I=0$,
which implies that $\bar\I\subseteq (1-E_\K)\W$. Thus we obtain
$\bar\I=(1-E_\K)\W$. From this it immediately follows that
$\bar\N=\K^\orth$ because $(1-E_\K)\H=\bar\I\H\subseteq\bar\N$ and
$E_\K\N=\{0\}$.
\QED
\end{proof}

Hence if $\N$ is closed, $\K$ is the unique physical acausal subspace
in the theory from \Cor{cor:SeparationOfNullSector}.  Further
$\bar\I=(1-E_\K)\W$ indicates that the finite-sum condition in the
definition of $\I$ cannot be removed in general.

\subsection{Type of $\VNA$-algebra $\C$ and irreducible physical subalgebras}

In general factors of $\VNA$-algebras are classified into the following
three types\CITE{Schwartz.J1967B}:
\begin{itemize}
\item[Type I:] A factor containing a non-zero minimum projection operator.
\item[Type II:] A factor which contains no non-zero minimum projection operator
but contains a finite non-zero projection operator.
\item[Type III:] A factor which contains no non-zero minimum projection
operator
and no finite non-zero projection operator.
\end{itemize}
Here a projection operator $E$ in a factor $\A$ is minimum if there
exists no projection operator $F$ in $\A$ such that $E\ge F$ and
$E\not=F$.  Further a projection operator $E\in\A$ is finite if for
any projection operator $F\in\A$ such that $E\ge F$ and $E\not=F$
there exists no partial isometry $V\in\A$ such that $V^*V=E$ and
$VV^*=F$.

In this paper we mainly consider the cases for which physical
subalgebras are of type I for the following reason. First recall that
in our formalism a physical subalgebra $\B$ was introduced as a
subalgebra of $\C$ which corresponds to all the functions on the
intersection of an acausal submanifold and the constraint submanifold
in the classical theory. Hence it is natural to require that $\B$
contains an enough amount of operators to separate vectors in a physical
acausal subspace $\L$ on which $\B$ acts, i.e., for any $x, y\in\L$
such that $x\not=ky$($k\in\CF$) and $||x||=||y||$ there exists an
operator $A\in\B$ such that $(x,Ax)\not=(y,Ay)$. From the following
proposition this requirement is equivalent to the requirement that
$\B$ is represented on $\L$ irreducibly.

\begin{proposition}
For a factor $\A$ and a closed subspace $\L$ such that $\A\L\subseteq\L$
$\A$ separates vectors in $\L$ if and only if the representation of $\A$
on $\L$ is irreducible. Further the latter condition holds if and only
if $\A_\L=\B(\L)$ where $\A_\L$ is the closure in $\B(\L)$ of the
restriction of $\A$ on $\L$.
\end{proposition}
\begin{proof}
Let $\A_\L'$ be the commutant of $\A_\L$ in $\B(\L)$. Then since $\A$
is a factor, $\A_\L\cap\A_\L'=\CF$. Suppose that $\A_\L\not=\B(\L)$.
Then $\A_\L'$ is not trivial because otherwise $\A_\L=\A_\L''=\B(\L)$.
Let $E$ be an orthogonal projection operator in $\A_\L'$ such that
$E\not=0,1$. Then there exists a vector $x\in\L$ such that
$y:=Ex\not=0$ and $z:=(1-E)x\not=0$. Clearly $(y,Az)$=0 for any
$A\in\A_\L$. Hence $(y\pm z, A(y\pm z))=(y,Ay)+(z,Az)$, or
$(y+z,A(y+z))=(y-z,A(y-z))$.  Since $||y-z||=||y+z||\not=0$ and $y-z$
and $y+z$ is not parallel, this implies that $\A$ does not separate
vectors in $\L$.  Conversely if $\A_\L=\B(\L)$, $\A$ clearly separates
vectors in $\L$.

Finally $\A$ is reducible on $\L$ if and only if there exists an
orthogonal projection operator $E$ in $\B(\L)$ which commutes with
$\A_\L$, i.e., $E\in\A_\L'$. Since $\A_\L\cap\A_\L'=\CF$, this is
equivalent to $\A_\L\not=\B(\L)$. \QED
\end{proof}

If $\B$ is a factor of type II(or III) and leaves $\L$ invariant, the
instant operator $E$ of $\L$ commutes with $\B$, hence $E\in\B'$.
However, since $\B'$ is also a factor of type II(III), it contains a
nontrivial orthogonal projection operator $F$ such that $E\ge F$ and
$E\not=F$. This implies that $\B$ is reducible on $\L$. Hence $\B$
cannot separate vectors in $\L$ from the proposition. In fact the
range of any orthogonal projection operator contained in a factor of
type II or III is always infinite-dimensional.

On the other hand if $\B$ is a factor of type I, there always exists a
closed subspace $\L$ on which $\B$ is represented irreducibly.  To see
this, let us first note that type I factors on $\H$ are subdivided
into the spatial isomorphism classes
$I_{m,n}$($m,n=1,\cdots,+\infty$), and for a factor $\B$ of type $I_{m,n}$
there exists a $m$-dimensional and $n$-dimensional Hilbert spaces,
$\hat H_m$ and $\hat H_n$ such that
\Beqr
&& \H \cong \hat L_2(\sigma(\Lambda),\mu)\bar\otimes\H_m\bar\otimes\hat \H_n,\\
&& \B \cong 1\otimes1\otimes \B(\hat \H_n).
\Eeqr
Hence for any non-zero vector $u\in\hat\H_m$, $\B$ is represented on the
subspace $u\otimes\hat\H_n$ irreducibly. Of course this does not imply
that there exist irreducible physical acausal subspaces for $\B$. We
will return to this central problem below after explaining an
additional technical assumption.

Here note that for type I cases $\W$ coincides with $\Z$ if and only
if $m=1$. Since $\W=\Z$ is the condition for the constraint algebra
$\W$ is abelian, the system with $\C$ of type $I_{1,n}$ will be called
an \newterm{abelian totally constrained system}. For the other cases
including the type-II and the type-III cases the constraint algebra
is non-abelian.

\subsection{Additional condition}

Since $\C=\W'$, $\C\N=\N$. Hence for any unitary operator $U$ in $\C$,
if $\L$ is an acausal subspace, $U\L$ is also an acausal subspace.
{}From this it follows that $\L$ is a physical acausal subspace for a
physical subalgebra $\B$, $U\L$ is a physical acausal subspace for the
physical subalgebra $U\B U^{-1}$. Taking account of the freedom of the
physical subalgebras, it is desirable if the same holds for any
unitary operator $U$ in $\Z'=\VNA(\C,\W)$ such that $U\B
U^{-1}\subseteq\C$. Unfortunately, however, it is not the case because
$U\N$ is not contained in $\N$ in general due to the finite-sum
condition in $\N$. In such a situation physical subalgebras are not
physically equivalent in general and we are obliged to classify them
into subclasses. However, such classification is unsatisfactory
because there is no good physical criterion to pick up one subclass.
Hence in the present paper we restrict the consideration to the cases
in which any $U\in\Z'$ such that $U\B U^{-1}\in\C$ maps $\N$ into itself.

In the case in which physical subalgebras are factors of type I we can
replace this restriction by a simple condition on $\I$ owing to the
following proposition.

\begin{proposition}
Let $\A$ be a factor of type I on a Hilbert space $\H$, and $U$ be
a unitary operator. Then $U\A U^{-1}=\A$ holds
if and only if there exists unitary operators $U_1\in\A$ and
$U_2\in\A'$ such that $U=U_1U_2$.
\end{proposition}
\begin{proof}
Since $\A$ is a factor of type I, it is spatially isomorphic to
$1\otimes\B(\hat\H_n)$ on $\hat\H_m\bar\otimes\hat\H_n$ for some positive
integer $m$ and $n$. Let $\{E_j\}$ and $\{F_k\}$ be the set of
mutually orthogonal 1-dimensional projection operators such that
$\sum_j E_j=1_m$ and $\sum_k F_k=1_n$, respectively. Suppose that $\A$
is invariant under $U$. Then since $U$ maps each minimum projection
operator to a minimum projection operator, there exists a unitary
operator $V$ on $\hat \H_n$ such that
$$
 U(1\otimes F_k)U^{-1}=1\otimes VF_kV^{-1}.
$$
Further for $A\in\A'$ and $B\in\A$,
$$
UAU^{-1}B=UA(U^{-1}BU)U^{-1}=U(U^{-1}BU)AU^{-1}=BUAU^{-1}
$$
since $U^{-1}BU\in\A$ from the assumption. This implies that
$U\A'U^{-1}=\A'$.  Hence from the same argument on the factor $\A'$ it
follows that there exists a unitary operator $W$ on $\hat\H_n$ such that
$$
U(E_j\otimes 1)U^{-1}=WE_j W^{-1}\otimes 1.
$$
Hence if we put $U_1=1\otimes V$ and $U_2=W\otimes1$, we obtain
$$
U(E_j\otimes F_k)U^{-1}=(U_1U_2)(E_j\otimes F_k)(U_1U_2)^{-1}.
$$
Since $\{E_j\otimes F_k\}$ spans $\B(\H)$, this implies that
$U=cU_1U_2$ where $c$ is a complex number such that $|c|=1$, which can
be absorbed into the definition of $U_1$ or $U_2$. The converse is
obvious. \QED
\end{proof}

When a physical subalgebra $\B$ is a factor of type I, for an
appropriate direct integral representation of $\H$,
\Beq
\H=\int\oplus \H(\lambda)d\mu(\lambda); \quad \H(\lambda)=\hat \H,
\Eeq
$\C$ is written in terms of a direct integral as
\Beq
\C\cong \int \oplus \C(\lambda)d\mu(\lambda); \quad
\C(\lambda)=\hat\B,
\Eeq
and $\B$ as
\Beq
\B\cong \SetDef{\int\oplus V(\lambda)A V(\lambda)^{-1} d\mu(\lambda)}
{A\in\hat\B},
\Eeq
where $\hat\B$ is a factor of type I on $\hat\H$, and $V(\lambda)$
is a measurable family of unitary operators on $\hat \H$ such that
$V(\lambda)\hat \B V(\lambda)^{-1}=\hat \B$. Hence for a decomposable unitary
operator $U$,
\Beq
U\cong \int\oplus U(\lambda)d\mu(\lambda),
\Eeq
$U\B U^{-1}$ is contained in $\C$ if and only if
$U(\lambda)V(\lambda)\hat\B V(\lambda)^{-1}U(\lambda)^{-1}=\hat\B$. Hence
from the proposition it follows that there exists families of unitary
operators $U_1(\lambda)\in\hat\B$ and $U_2(\lambda)\in\hat\B'=\W(\lambda)$
such that $U(\lambda)=U_1(\lambda)U_2(\lambda)$(and corresponding
operators for $V(\lambda)$). $U_1(\lambda)$ and $U_2(\lambda)$ can be
chosen as measurable families of operators from the following lemma.

\begin{lemma}
Let $V_j(\lambda)$($j=1,\cdots,\infty$) be a family of functions defined
on a common domain $D\subseteq\RF$ such that
for any positive integer $j$ there exists a set $S_j$ of positive integers
such that $j\in S$ and $\sum_{k\in S_j}|V_k|^2=1$. Then there exists
a phase function $e^{i\theta(\lambda)}$ such that $V_j(\lambda)
e^{i\theta(\lambda)}$ becomes a $\mu$-measurable function for all $j$
if and only if $\bar V_j(\lambda) V_k(\lambda)$ is Borel measurable for
all $j$ and $k$.
\end{lemma}
\begin{proof}
Assume that $\bar V_j(\lambda) V_k(\lambda)$ are Borel measurable. Then
$$
|V_k(\lambda)|^2 = \sum_{l\in S_j}|V_l(\lambda)V_k(\lambda)|^2
$$
is Borel measurable. Hence if we put %
$$
V_k(\lambda)=A_k(\lambda)e^{i\theta_k(\lambda)}\quad(A_k(\lambda)\ge0,
\theta_k(\lambda)=0
\quad\r{if}\quad A_k(\lambda)=0),
$$
$A_k(\lambda)$ is Borel measurable. Let $D_j$ be the set
of $\lambda$ for which $A_j(\lambda)\not=0$. Then $D_j$
are Borel measurable sets and $\cup_j D_j=D$. Hence there exists a division
$E_n$ of $D$ such that each $E_n$ is Borel measurable and for each $n$
there exists a positive integer $J(n)$ such that $A_{J(n)}>0$ on $E_n$.
Let us define $\theta(\lambda)$ by
$$
\theta(\lambda)=-\theta_{J(n)}(\lambda) \quad\r{for}\quad \lambda\in E_n.
$$
Then since $\theta_j(\lambda)-\theta_k(\lambda)$ is Borel measurable for any
$j,k$,  $\theta_j(\lambda)+\theta(\lambda)$ is Borel measurable for any $j$,
which implies that $V_j(\lambda)e^{i\theta(\lambda)}$ is
Borel measurable. The converse is obvious.
\QED
\end{proof}

Thus $U_1(\lambda)$ and $U_2(\lambda)$ define unitary operators
$U_1\in\C$ and $U_2\in\W$, respectively, and $U$ is written as
$U=U_1U_2$. Therefore, noting that the linear combinations of unitary
operators are dense in $\W$,
the invariance of $\N$ by $U$ generally requires that
\Beq
\W\I=\I.
\label{AdditionalCondition}\Eeq
On the basis of this observation we assume that condition
\Eq{AdditionalCondition} is satisfied from now on, and denote
the set of unitary operators written as a product of a unitary
operator in $\C$ and one in $\W$ by $\U_\C$.

Here note that in stead of regarding \Eq{AdditionalCondition} as a
condition on the quantum totally constrained system, we can replace
$\I$ by $\J$ defined by
\Beq
\J:=\W\I,
\Eeq
and the linear null manifold $\N$ by $\J\H$. Then the condition
(\ref{AdditionalCondition}) is automatically satisfied, and all the
statements and the propositions given before this subsection hold
without change.

\subsection{Quantum dynamics of type I systems}

Now we prove that quantum dynamics of totally constrained systems
is consistent in the sense that the causal mapping is always conformal
if the $\VNA$-algebra $\C$ is of type I and the additional condition
(\ref{AdditionalCondition}) is satisfied. In this section we fix
the spatial isomorphism
\Beqr
&& \H \cong L_2(\Lambda,\mu)\bar\otimes \hat\H_{m,n};\quad
\hat\H_{m,n}=\hat\H_m\bar\otimes\hat\H_n,\\
&& \C \cong L_\infty(\Lambda,\mu)\bar\otimes \hat\B;\quad
\hat\B=1\otimes\B(\hat\H_n),\\
&& \W \cong L_\infty(\Lambda,\mu)\bar\otimes \hat\W;\quad
\hat\W=\B(\hat\H_m)\otimes1.
\Eeqr
Further we denote the physical subalgebra corresponding to $1\otimes\hat\B$
by this fixed isomorphism by $\B_0$:
\Beq
\B_0\cong 1\otimes\hat\B.
\Eeq

We also often use the direct integral expressions for vectors
and operators. The isomorphism between the original Hilbert space
and algebras and the corresponding direct integral representations
is also understood to be fixed. In particular we denote the
family of operators on $\hat\H_m$ corresponding to the bounded
constraint operators as
\Beq
H_\alpha\cong \int_{\sigma(\Lambda)}\oplus \hat H_\alpha(\lambda)\otimes1
d\mu(\lambda),
\Eeq
and the subspace of $L_2(\sigma(\Lambda),\mu)\bar\otimes\hat\H_m
\cong L_2(\sigma(\Lambda),\mu;\hat\H_m)$ finitely generated by $\hat H_\alpha$
by $\hat \N$:
\Beq
\hat \N := \SetDef{\sum_\alpha \hat H_\alpha v_\alpha(\r{finite\ sum})}
{v_\alpha\in L_2(\sigma(\Lambda),\mu)\bar\otimes\hat\H_m}.
\Eeq
Here note that from the structure of algebras above the operation of $\W$
on $\N$ naturally induces an operation of $\W$ on $\hat\N$, and for this
operation the condition (\ref{AdditionalCondition}) yields
\Beq
\W\hat\N=\N.
\Eeq

First we prove the following proposition which clarifies the existence
and the distribution of irreducible physical acausal subspaces.

\begin{proposition}\label{proposition:StructureOfIPAS}
If the condition $\W\I=\I$ is satisfied, the structure of physical
acausal subspaces are described as follows:
\begin{itemize}
\item[1)] $\L_0=\overline{\B_0 x}$ yields an irreducible physical acausal
subspace if and only if $x\in\H$ is expressed as $x\cong z\otimes y$
in terms of $y\in\hat\H_n$ and $z\in L_2(\sigma(\Lambda),\mu)\bar\otimes
\hat\H_m$ such that $z\not\in\hat\N$. In that case $\L_0$ is written as
$\L_0\cong z\otimes\hat\H_n$.
\item[2)] A closed subspace $\L$ is an irreducible physical subspace
if and only if it is expressed as $\L=U\L_0$ in terms of a unitary
operator $U\in\U_\C$ and an irreducible physical acausal subspace $\L_0$
corresponding to the physical subalgebra $\B_0$ in 1).  Further its
physical subalgebra is given by $\B=U\B_0 U^{-1}$.
\end{itemize}
\end{proposition}
\begin{proof}
\noindent
1) In terms of an orthonormal basis $e_j$ of $\hat\H_n$ and a set of
vectors $v_j\in L_2(\sigma(\Lambda),\mu)\bar\otimes\hat\H_m$, $x\in\H$ is
written as
$$
x=\sum_j v_j\otimes e_j.
$$
Since $\B(\hat\H_n)$ contains the orthogonal projection operator $\hat E_j$
onto the one-dimensional subspace spanned by $e_j$ for any $j$,
$E_j\cong 1\otimes \hat E_j \in\B_0$, and
$E_j x= v_j\otimes e_j\in \overline{\B_0 x}$. Hence
$$
\B_0 E_j x =v_j\otimes \B(\hat\H_n)e_j=v_j\otimes\hat\H_n\subseteq
\overline{\B_0 x}.
$$
{}From this it follows that
$$
\L_0=\overline{\B_0 x}=Q\bar\otimes\hat\H_n,
$$
where $Q$ is the closure of the linear space spanned by $\{v_j\}$ in
$L_2(\sigma(\Lambda),\mu)\bar\otimes\hat\H_m$. Clearly $Q$ should be
one-dimensional in order that $\L_0$ is irreducible with respect to
$\B_0$. This implies that there should exist a set of constants $c_j\in\CF$ and
$z\in L_2(\sigma(\Lambda),\mu)\bar\otimes\hat\H_m$ such that
$v_j=c_j z$. Hence in terms of $y=\sum_j c_j e_j\in\hat\H_n$, $x$ is written as
$$
x=z\otimes y.
$$
Here $x\in\N$ if and only if $z\in\hat\N$. Hence $\L_0$ is acausal if
and only if $z\not\in\hat\N$.

\noindent
2) Let $\L$ be an irreducible physical acausal subspace invariant under
a physical subalgebra $\B=U\B_0 U^{-1}$ where $U\in\U_\C$:
$$
\B\L=U\B_0 U^{-1}\L=\L.
$$
Then since the transformation $U$ preserves the acausality and the
irreducibility, this condition is equivalent to the condition that
$U^{-1}\L$ is an irreducible physical acausal subspace for $\B_0$.
Hence $\L$ is written as $\L=U\L_0$ in terms of an irreducible physical
acausal subspace $\L_0$ for $\B_0$.
\QED
\end{proof}

This proposition shows that there may exist acausal vectors which are
not contained in any irreducible physical acausal subspace in general.
However, for the abelian constrained systems such pathology does not
occur.

\begin{corollary}
If $\W=\Z$, every vector in $\H-\N$ is contained in some irreducible
physical acausal subspace.
\end{corollary}
\begin{proof}
First note that if $\W=\Z$, $\H_m$ is one-dimensional and is absorbed
into $\L_2(\sigma(\Lambda),\mu)$.
Let $x$ be a vector in $\H-\N$. $x$ is expressed in the direct integral
representation as
$$
x\cong \int_{\sigma(\Lambda)}\oplus x(\lambda) d\mu(\lambda).
$$
Since $x(\lambda)$ is $\mu$-measurable, there exists a $\mu$-measurable
family of unitary operators $U(\lambda)\in\B(\hat\H_n)$, a measurable
function $\phi(\lambda)=||x(\lambda)||$, and a constant vector
$y\in\hat\H_n$ such that $x(\lambda)=\phi(\lambda)U(\lambda)y$.
Since $\U_\C=\Z'=\C$ now, $U$ belongs to $\U_\C$, and $\phi\otimes y$
does not belong to $\N$. Hence from the proposition, $\L=U(\phi\otimes
\hat\H_n)$ is an irreducible physical acausal subspace which contains
$x$ and invariant under $U\B_0 U^{-1}$.
\QED
\end{proof}

Now we prove the main theorem.

\begin{theorem}\label{theorem:conformality1}
If the $\VNA$-algebra $\C$ is of type I and the condition $\W\I=\I$ is
satisfied, the causal mapping $\Theta$ between arbitrary pair of
irreducible physical acausal subspaces $\L_1$ and $\L_2$,
$$
\Theta:\L_1\cap D(\L_2) \maps \L_2\cap D(\L_1)
$$
is conformal. In particular if $\L_1\cap\L_2\not=\{0\}$, or if there exists
a unitary transformation $U\in\U_\C$ such that $\L_2=U\L_1$,
$\Theta$ is isometric.
\end{theorem}
\begin{proof}
Since any unitary transformation $U$ in $\U_\C$ preserves the physical
acausality and conformality, we can assume without lose of generality that
$\L_1$ and $\L_2$ are expressed in terms of a unitary operator $U\in\U_\C$
and vectors $z_1,z_2\in L_2(\sigma(\Lambda),\mu)\bar\otimes\hat\H_m$ such that
$z_1,z_2\not\in\hat\N$ as
$$
\L_1\cong z_1\otimes \hat\H_n, \quad \L_2\cong U(z_2\otimes \hat\H_n).
$$
Let $x$ be a vector in $\L_1\cap D(\L_2)$. Then $x$ and  $\Theta(x)$
are expressed in terms of  direct integrals as
\begin{eqnarray*}
&&x=\int\oplus z_1(\lambda)\otimes y d\mu(\lambda),\\
&&\Theta(x)=\int\oplus W(\lambda)z_2(\lambda)\otimes V(\lambda)\hat\Theta(y)
d\mu(\lambda),
\end{eqnarray*}
where $y$ and $\hat\Theta(y)$ are non-zero vectors in $\hat\H_n$, and
$W(\lambda)\in\B(\hat\H_m)$ and $V(\lambda)\in\B(\hat\H_n)$ are
measurable families of unitary operators such that $U=VW$. Hence the condition
$\Theta(x)-x\in\N$ is expressed as
$$
W(\lambda)z_2(\lambda)\otimes V(\lambda)\hat\Theta(y)
-z_1(\lambda)\otimes y = \sum_\alpha \hat H_\alpha(\lambda)u_\alpha(\lambda),
$$
where $u_\alpha(\lambda)$ are a finite number of measurable families of
vectors in $\hat\H_m\bar\otimes\hat\H_n$.
This equation is equivalent to the following two conditions:
\Beqr
&& W(\lambda) z_2(\lambda)k(\lambda)- z_1(\lambda) ||y||^2 \in \hat \N,
\label{tmp:causality1}\\
&& W(\lambda)z_2(\lambda)||\hat\Theta(y)||^2-z_1(\lambda) \bar k(\lambda)\in
\hat\N,\label{tmp:causality2}
\Eeqr
where $k(\lambda):=(y,V(\lambda)\hat\Theta(y))$. Similarly for another
$x'\in\L_1\cap D(\L_2)$, by denoting the corresponding quantities
by symbols with a prime, we obtain
\Beqr
&& W(\lambda) z_2(\lambda)k'(\lambda)- z_1(\lambda) ||y'||^2 \in \hat \N,
\label{tmp:causality3}\\
&& W(\lambda)z_2(\lambda)||\hat\Theta(y')||^2-z_1(\lambda) \bar k'(\lambda)\in
\hat\N. \label{tmp:causality4}
\Eeqr
The combination of these equations,
$||\hat\Theta(y')||^2\times$\Eq{tmp:causality1}
$+k'(\lambda)\times$\Eq{tmp:causality2}\\
$-||\hat\Theta(y)||^2\times$\Eq{tmp:causality3}
$-k(\lambda)\times$\Eq{tmp:causality4}, yields
$$
\left(||y'||^2||\hat\Theta(y)||^2-||y||^2||\hat\Theta(y')||^2
+k(\lambda)\bar k(\lambda')-\bar k(\lambda)k(\lambda')\right)z_1(\lambda)
\in\hat\N.
$$
Since $k(\lambda)\bar k(\lambda')-\bar k(\lambda)k(\lambda')$ is pure
imaginary, from this equation and the condition $z_1\not\in\hat\N$
it follows that
$$
||y'||^2||\hat\Theta(y)||^2-||y||^2||\hat\Theta(y')||^2=0,
$$
or equivalently
$$
{||\hat\Theta(y)|| \over||y||}={||\hat\Theta(y')|| \over||y'||}.
$$
Hence we obtain for any pair of non-zero vectors $x, x'\in\L_1\cap D(\L_2)$
$$
{||\Theta(x')|| \over||x'||}={||z_2|| ||\hat\Theta(y')|| \over||z_1|| ||y'||}
={||z_2|| ||\hat\Theta(y)|| \over||z_1|| ||y||}={||\Theta(x)|| \over||x||}.
$$
This shows that $\Theta$ is conformal.

If $\L_1\cap\L_2\not=\{0\}$, the conformal factor should be unity since
there exists a vector $x\in\L_1\cap D(\L_2)$ such that $\Theta(x)=x$.
Further if $\L_2=U\L_1$ for some $U\in\U_\C$, $z_1=z_2$. Hence
\Eqs{tmp:causality1}{tmp:causality2} are written as
\begin{eqnarray*}
&&\left(W(\lambda)k(\lambda)-||y||^2\right)z_1(\lambda) \in \hat\N,\\
&&\left(||\hat\Theta(y)||^2-W^{-1}(\lambda)k(\lambda)\right)z_1(\lambda)
\in \hat\N.
\end{eqnarray*}
However, since $||W(\lambda)|| |d(\lambda)|\le ||y|| ||\hat\Theta(y)||$,
if $||y||\not=||\hat\Theta(y)||$, $z_1$ must belong to $\hat\N$, contradicting
the assumption. Hence $||\Theta(x)||/||x||=1$.
\QED
\end{proof}

If we restrict the argument to irreducible acausal subspaces with the same
physical subalgebra, we can show the following stronger result.

\begin{theorem}\label{theorem:conformality2}
For the type-I totally constrained system satisfying the condition
$\W\I=\I$, let $\L_1$ and $\L_2$ be physical acausal subspaces invariant
under the same physical subalgebra $\B$ such that $\L_1$ is irreducible
and $\L_1+\L_2$ is not acausal. Then $\L_2':=\L_2\cap D(\L_1)$ is
an irreducible physical acausal subspace for $\B$, and the causal
mapping $\Theta$ is conformal and bijective from $\L_1$ onto $\L_2'$.
\end{theorem}
\begin{proof}
In terms of a unitary transformation in $\U_\C$ we can reduce the statement
to the case in which $\B=\B_0$ and
$$
\L_1 \cong z_1\otimes\hat\H_n.
$$
Since $\L_1+\L_2$ is not acausal, there exists a vector $x_0\in\L_1$
such that $\Theta(x_0)\in\L_2$. Then for any vector $x\in\L_1$
there exists $A\in\B_0$ such that $x=Ax_0$, and
$$
A\Theta(x_0)-Ax_0=A\Theta(x_0)-x \in\N.
$$
Since $\L_2$ is invariant under $\B_0$, this implies that $x\in\domain{\Theta}$
and $\Theta(x)=A\Theta(x_0)$. Hence $\domain{\Theta}=\L_1$ and
$$
\Theta(Ax)=A\Theta(x) \quad \forall A\in\B_0, \quad \forall x\in\L_1.
$$
However, since $\Theta$ is conformal from the previous theorem,
$\L_2'=\L_2\cap D(\L_1)=\Theta(\L_1)$ is closed, and obviously
invariant under $\B_0$. Further since $\Theta$ is bijective and commutes
with the operation of $\B_0$, the irreducibility of $\L_2'$ follows from
that of $\L_1$.
\QED
\end{proof}

\begin{corollary}\label{corollary:conformality2}
Let $\L_1$ be an irreducible physical acausal subspace with respect to
$\B$ and $x$ be a vector in $\in D(\L_1)$. Then under the same condition
on $\C$ as in the theorem, $D(\L_1)=D(\L_2)$ if $\L_2:=\overline{\B x}$
is acausal.
\end{corollary}

Now we comment on the implications of these theorems.
First for the abelian totally constrained system, when a state vector
$u_0\not\in\N$ is given by observation, we can always find
an irreducible physical acausal subspace $\L_0$ containing $u_0$.
Then the normalized initial value of the relative probability
amplitude $\Psi$ on $\L_0$ is determined by
\Eq{InitialValueProblemForPsi}, which in turn determines $\Psi$ on
$D(\L_0)$ by the weak hamiltonian constraint. Here the choice $\L_0$
is not unique, but \Theorem{theorem:conformality1} guarantees that
the values of $\Psi$ determined from different initial irreducible
physical acausal subspaces $\L_0$ and $\L_0'$ coincides with each other on the
intersection of the domain of dependence since
\Beq
\Psi|_{\L_0'}(\Theta(u))={(u_0,\Theta(u))\over||u_0||^2}
={(\Theta(u_0),\Theta(u))\over||\Theta(u_0)||^2}
={(u_0,u)\over||u_0||^2}=\Psi|_{\L_0}(u).
\Eeq
Thus $\Psi$ is in effect determined on a subset
\Beq
D(u_0):=\bigcap_{\begin{array}{l}U\in\C \\ Uu_0=u_0\end{array}}D(U\L_0).
\Eeq

This limitation of the domain of $\Psi$ restricts predictions strongly.
First for state vectors which are not contained in  $D(u_0)$ we can say
nothing about its probability. Second, though for any  vector $u$ in
$D(u_0)$ there exists an irreducible physical acausal subspace $\L$
containing $u$, it may not be fully contained in
$D(u_0)$ in general. In such a case we are obliged to restrict the
prediction on state vectors contained in a closed subspace of
$\overline{\L\cap D(u_0)}$. Owing the \Theorem{theorem:conformality1}
$\Psi$ gives a normalizable probability on such a subspace.

One reason why this kind of pathology occurs in quantum theory is that
the instant operators have much larger freedom than the instant functions
in classical theory. In classical theory the instant functions should be
chosen so that the corresponding acausal submanifolds are smooth.
On the other hand, in quantum theory, the instant operators are not
restricted to those which correspond to such smooth instant functions
in classical theory.  Intuitively speaking, the category of functions
are extended to that of measurable functions. Meanwhile, since we have
not taken the closure of the linear null manifold, we are requiring
some kind of smoothness in the dynamical correspondence between acausal
subspaces. Hence for two acausal subspaces for which the corresponding
instant operators do not have good smoothness relatively, they cannot be
dynamically related. For example in a single constraint system, for two
irreducible acausal subspaces given by $\L_1\cong z_1\otimes\hat\H_n$
and $\L_2\cong z_2\otimes\hat\H_n$, the causal mapping becomes bijective
if $z_1(\lambda)$ and $z_2(\lambda)$ are both smooth function of $\lambda$.
However, if one of them is smooth and the other is just measurable,
they become acausal with each other in general. Further if we consider the
freedom of the unitary transformation of them by $U$ in $\U_\C$,
they may also become partially causal occur depending on the smoothness of
$U(\lambda)$.  Therefore the incomplete dynamical correlation
in the full state space is inevitable as long as we allow states
corresponding to eigenstates of all possible instant operators or
time variables.

Another source of the pathology is related with the good time variables
in the classical theory. As shown in the previous paper\CITE{Kodama.H1995},
for a single constraint system, the conservative measure on an acausal
submanifold $\Sigma$ coincides with the natural measure induced from
the canonical volume form of the phase space and the instant function $f$
specifying $\Sigma$ only when $f$ is a good time variable, i.e., $\{f,h\}$
is a constant of motion. Though this does not restrict a single acausal
submanifold, it restricts the possible pairs of acausal submanifolds which
correspond to two different values of a good time variable. Since
the measures or the inner products of acausal subspaces are induced
from the inner product of the whole Hilbert space in quantum theory,
instant operators correspond to the good time variables in the
classical theory. Hence it is naturally expected that the unitarity
is violated for pairs of acausal subspaces whose instant operators
do not correspond to the same good time variable.

Taking account of this observation, one natural way to avoid these
complications in dynamics is to restrict the prediction on
such irreducible physical acausal subspaces that are contained in
$D(\L_0)$ for some irreducible physical acausal subspace $\L_0$ passing
through $u_0$. \Theorem{theorem:conformality2} and
\Cor{corollary:conformality2} gives the simplest family of such
irreducible physical acausal subspaces. In reality there exist a huge
class of such irreducible physical acausal subspaces, although it
is difficult to give a simple characterization of them in general.
For example in the single constraint system $(\H,h)$,
for $\L_1\cong z_1\otimes \hat\H_n$ and any vector
$\hat n\in\hat\N$, $\L_2\cong U((z_1+\hat n)\otimes\hat\H_n)$ satisfies
the required condition if $U(\lambda)$ depends on $\lambda$ smoothly.
This example can be easily extended to the case of multiple abelian
constraint systems with small modification.  In fact, in systems corresponding
to ordinary quantum mechanical system, only such a special class of
irreducible physical acausal subspaces are relevant as will be discussed
in the next subsection. Further, if we consider only the consistent pairs
of irreducible physical acausal subspaces, we can describe the statistical
dynamics of the system in mixed dynamical states consistently as well.

For the non-abelian totally constraint systems, the situation is the
same as for the abelian case if they are of type I, except for one point.
It is the poverty of irreducible physical acausal subspaces. As
\Prop{proposition:StructureOfIPAS} shows, a state vector is contained
in an irreducible physical acausal subspace if and only if it is isomorphic
to a vector of the form $U(z\otimes y)$ with a unitary operator $U\in\U_\C$.
Hence in the non-abelian case for which $\U_\C$ does not cover all the
unitary operators in $\Z'$, irreducible physical acausal subspaces
pass just a proper subset $\SS$ of $\H-\N$. This is a quite embarrassing
situation since if the given state vector $u_0$ is not contained in $\SS$,
we can make no consistent dynamical prediction in our formalism.

Though it is not clear whether this pathology is serious in
practical situations,  it is possible to make the set of the allowed
state vectors much larger to the point so that it can be regarded as
maximal by slightly modifying the formalism within the general framework.
To see this, first note that for any vector $x\in\H$ there exists a
unitary operator $U$ in $\Z'$ such that $x\cong U(z\times y)$.
Let us replace the additional condition \Eq{AdditionalCondition} by
\Beq
\Z'\I=\I
\label{AdditionalCondition2}\Eeq
or replace $\I$ by $\Z'\I$. This enlarges the linear null manifold and
the causally related states. In particular, since $\Z'\N=\N$ now, $U$ above
transform the irreducible physical acausal subspace $\L_0\cong z\otimes\hat
\H_n$ to an acausal subspace $\L$ through $x$, which is invariant and
irreducible with respect to $\A:=U\B_0 U^{-1}$. $\A\cap\W$ may not be
trivial any longer, but $\L=\overline{\A x}$ is still acausal.  Further
by inspecting the proof of \Theorem{theorem:conformality1} one easily check
that the causal mapping between two acausal subspaces isomorphic to
$U_1(z_1\times\hat\H_n)$ and $U_2(z_2\times\hat\H_n)$ is conformal at
least if $U_1^{-1}U_2\in\U_\C$ even when $U_1,U_2\in\Z'$. Of course
$\A=U\B_0 U^{-1}$ may not be contained in $\C$. If one wants to restrict
the algebra to a subset of $\C$, one may replace $\A$ by the $\VNA$-subalgebra
$\A_1=\A\cap \C$, which is a factor and $\A_1\cap\Z=\CF$. However, then
$\L_1$ may become reducible with respect it.

The condition (\ref{AdditionalCondition2}) is rather strong. It actually
requires that the norms of $H_n(\lambda)$ vanish simultaneously at least
at some value of $\lambda$, roughly speaking, and the dynamics is essentially
determined by an abelian set of constraint operators, as the following
proposition shows.

\begin{proposition}
If the condition $\Z'\I=\I$ holds, there exists a
family of self-adjoint operators in $\Z$ such that
\begin{itemize}
\item[a)] $0\le H_1 \le H_2 \le \cdots$,
\item[b)] there exists $A_n\in\Z$ such that $H_n=H_{n+1}A_n$,
\item[c)] $\N=\bigcup_n H_n\H$.
\end{itemize}
\end{proposition}
\begin{proof}
The proof is a little bit long. So we divide it into three steps.

\noindent
1) Let $H$ be a non-negative bounded self-adjoint operator on a Hilbert
space $\hat\H$, and let its spectral decomposition be
$$
H=\int_0^\infty \xi dE(\xi).
$$
Further let $P_n(t)$ be a sequence of polynomials which converge
to zero locally uniformly on $1\le t\le C$ and to unity locally
uniformly on $0\le t<1$ where $C$ is a sufficiently large positive
constant. Then there exists sequences $\delta_n$ and $\epsilon_n$
of positive constants such that $\lim_{n\tend+\infty}\delta_n, \epsilon_n
=0$ and
\begin{eqnarray*}
&&|P_n(t)| \le \epsilon_n \quad 1\le t\le C,\\
&&|P_n(t)-1| \le \epsilon_n \quad 0\le t < 1-\delta_n.
\end{eqnarray*}
Hence for a positive constant $a$ and $x\in\hat\H$ we obtain the estimate
$$
||P_n\left(H/ a-E(\xi<a)\right)x||^2 \le \epsilon_n||x||^2
+ (x,E(a(1-\delta_n)\le\xi<a)x).
$$
The right-hand side of this equation tends to zero in the limit $n\tend\infty$.
This implies that
$$
\SOT-\lim_{n\tend\infty} P_n({H\over a})=E(\xi<a).
$$

Now let $H(\lambda)$ be a $\mu$-measurable family of self-adjoint uniformly
bounded non-negative operators on $\hat\H$, i.e., $H(\lambda)\in
L_\infty(\Lambda,\mu;\hat\H)$, and $\phi(\lambda)$ be a positive uniformly
bounded $\mu$-measurable function. Then $P_n(H(\lambda)/\phi(\lambda))$
is a $\mu$-measurable family of operators and from the above result
$$
\SOT\lim_{n\tend\infty}P_n({H(\lambda)\over\phi(\lambda)})
=E_\lambda(\xi<\phi(\lambda))\quad (\mu-\r{a.e.}),
$$
where $E_\lambda(\xi)$ is the spectral measure for $H(\lambda)$.
Hence $E_\lambda(\xi<\phi(\lambda))$ and $E_\lambda(\xi\ge\phi(\lambda))$
are $\mu$-measurable families of orthogonal projection operators.

\noindent
2) Let $H(\lambda)$ be the same family of operators in 1). Then for a
orthonormal basis $\{e_j\}$ of $\hat\H$, there exists a set of
Borel step functions, $H_{(n)jk}(\lambda)$, and a sequence $\epsilon_n$
of positive constants tending to zero, such that $\bar H_{(n)jk}(\lambda)
=H_{(n)kj}(\lambda)$ and
$$
|H_{jk}(\lambda)-H_{(n)jk}(\lambda)|\le \epsilon_n,
$$
where $H_{jk}(\lambda):=(e_j,H(\lambda)e_k)$. If we define a $\mu$-measurable
family of operators $H_{(n)}(\lambda)$ by
$$
H_{(n)}x=\sum_{j,k} e_j H_{(n)jk}(e_k,x),
$$
we obtain the estimate
$$
|(x,(H(\lambda)-H_{(n)}(\lambda))y)|\le\epsilon_n||x||||y||.
$$
Hence $H_{(n)}(\lambda)$ turns out to be a family of bounded self-adjoint
operators, and converges to $H(\lambda)$ uniformly with respect to $\lambda$:
$$
\lim_{n\tend\infty}||H(\lambda)-H_{(n)}(\lambda)||=0\quad
(\r{uniformly\ w.r.t.}\quad\lambda).
$$
Therefore from Weyl's theorem\CITE{Akiezer.N&Glazman1966B} we obtain
$$
|\phi(\lambda)-\phi_n(\lambda)|\le||H(\lambda)-H_{(n)}(\lambda)||
\tend 0\quad
(\r{uniformly\ w.r.t.}\quad\lambda),
$$
where $\phi(\lambda)$ and $\phi_n(\lambda)$ are the maximums of
the continuous spectra of $H(\lambda)$ and $H_{(n)}(\lambda)$,
respectively. Since $\phi_n(\lambda)$ is $\mu$-measurable from
the definition of $H_{(n)}(\lambda)$, this implies that
$\phi(\lambda)$ is also $\mu$-measurable.

\noindent
3) For a bounded constraint operator $H$ with the spectral decomposition
$$
H=\int H(\xi)dE(\xi),
$$
let $U$ be the unitary operator defined by $U=E(\xi\ge0)-E(\xi<0)$.
Then $H=|H|U$, and $HA\in\I$ can be written as $|H|UA$, and $|H|A=HUA\in\I$
for any $A\in\W$. Hence in the definition of $\I$ and $\N$, we can replace
$H_\alpha$ by $|H_\alpha|$. So from now on we assume that $H_\alpha\ge0$.

For the central decomposition of $H$,
$$
H=\int\oplus H(\lambda)d\mu(\lambda),
$$
let $\phi(\lambda)$ be the maximum value of the continuous spectrum of
$H(\lambda)$, and decompose $H(\lambda)$ as
$$
H(\lambda)=H_{(1)}(\lambda)+H_{(2)}(\lambda),
$$
where
$$
H_{(1)}:=H(\lambda)E_\lambda(\xi<2\phi(\lambda)).
$$
Then $\tilde H(\lambda):={\phi(\lambda)^{-1}}H_{(1)}(\lambda)$($=0$, for
$\lambda$ such that $\phi(\lambda)=0$) yields a measurable family of
self-adjoint operators uniformly bonded with respect to $\lambda$. Hence
we obtain
\begin{eqnarray*}
&H(\lambda)\W(\lambda)&=\phi(\lambda)\tilde H(\lambda)\W(\lambda)
+H_{(2)}(\lambda)\W(\lambda)\\
&&\subseteq \phi(\lambda)\W(\lambda)+ H_{(2)}(\lambda)\W(\lambda).
\end{eqnarray*}
On the other hand for the interval
$\Delta_\lambda:=[{1\over2}\phi(\lambda),2\phi(\lambda)]$,
\begin{eqnarray*}
&\phi(\lambda)E_\lambda(\Delta_\lambda)&
=\int_{\Delta_\lambda}\xi dE_\lambda(\xi)
\int_{\Delta_\lambda}{\phi(\lambda)\over\xi}dE_\lambda(\xi)\\
&&=H(\lambda)\int_{\Delta_\lambda}{\phi(\lambda)\over\xi}dE_\lambda(\xi)\\
&&\in H(\lambda)\W(\lambda).
\end{eqnarray*}
Let $\chi(\lambda)$ be the characteristic function for the set
$\sigma_+=\SetDef{\lambda}{\phi(\lambda)>0}$. Then it is a
$\mu$-measurable function. Since the projection operator
$E_\lambda(\Delta_\lambda)$ is $\mu$-measurable as a function of
$\lambda$ and its range is infinite-dimensional, there exists a
$\mu$-measurable family of operators
$A(\lambda),B(\lambda)\in\Z'(\lambda)$ such that $\chi(\lambda)
1=A(\lambda)E_\lambda(\Delta_\lambda)B(\lambda)$, hence
$$
\phi(\lambda) 1=A(\lambda)\phi(\lambda)E_\lambda(\Delta_\lambda)B(\lambda).
$$
However, since $\Z'\N=\N$, this implies that $\phi\H\subseteq\N$.
Hence we obtain
$$
\N=\sum_\alpha \phi_\alpha\H + \sum_\beta H_\beta\H,
$$
where $H_\beta$ is a self-adjoint operator such that the maximum of
its spectrum is a point spectrum of a finite multiplicity if $H_\beta(\lambda)
\not=0$.

Let $\psi_\beta(\lambda):=||H_\beta(\lambda)||$. Then $\psi_\beta(\lambda)$,
hence $E_\lambda(\xi\ge\psi_\beta(\lambda))$, is $\mu$-measurable and
an orthogonal projection operator onto a finite-dimensional subspace of
$\hat\H$ for each $\lambda$ if $\psi_\beta(\lambda)>0$. Therefore
there exists a $\mu$-measurable vector $x_0(\lambda)$ such that
$||x_0(\lambda)||=1$ and
$$
H_\beta(\lambda)x_0(\lambda)=\psi_\beta(\lambda)x_0(\lambda) \in\N.
$$
However,  since $\Z'\N=\N$, this implies that for any unitary $U\in\Z'$
and any $f(\lambda)\in L_2(\sigma(\Lambda),\mu)$,
$$
\psi_\beta(\lambda)f(\lambda)U(\lambda)x_0(\lambda) \in \N.
$$
Since the set of all vectors of the form $f(\lambda)U(\lambda)x_0(\lambda)$
coincides with $L_2(\sigma_+,\mu;\hat\H)$, it follows that
$\psi_\beta\H\subseteq\N$. On the other hand, from $||\psi_\beta(\lambda)^{-1}
H_\beta(\lambda)||\le1$, it follows that $H_\beta x\in \psi_\beta\H$.
Thus we obtain
$$
\N=\sum_\alpha\phi_\alpha \H + \sum_\beta\psi_\beta\H.
$$

Let us define the sequence of $\mu$-measurable self-adjoint operators
$H_n(\lambda)$ by
$$
H_n(\lambda):=\sup_{\alpha,\beta\le n}\left(\phi_\alpha(\lambda),
\psi_\beta(\lambda)\right).
$$
Then obviously $H_1\le H_2\le\cdots$, $A_n:=H_n/H_{n+1}\le1$ and
$$
\N=\cup_n H_n\H.
$$
\QED
\end{proof}

\begin{corollary}
For an abelian totally constraint system with a finite number of constraint
operators, there exists a bounded self-adjoint operator $\Lambda\in\Z$
and a Borel measurable function $\phi$ on $\RF$ such that
$\Z=\W=\VNA(\Lambda)$ and
$$
\N=\phi(\Lambda)\H.
$$
\end{corollary}

\subsection{Reduction}

When one wants to construct a quantum theory without constraints for a
totally constrained system, one usually first construct a classical
unconstrained canonical theory by solving the constraints under some
gauge-fixing condition and then quantize it. Here by gauge-fixing we
mean an appropriate choice of a family of acausal submanifolds or
instant functions and a reduction field in the phase
space\CITE{Kodama.H1995}.  In this method it is in general difficult
to compare two quantum theories corresponding to different
gauge-fixing because there exists no exact correspondence between the
observables within the quantum framework, though some of them such as
those related with symmetries may have good correspondence.

In contrast in our formalism we can fix the gauge or construct reduced
theories within the quantum framework and can compare them directly
because each state carries information on instants it belongs to and
the whole state space contains the states corresponding to all possible
instants.

The explicit reduction procedure in our formalism is given as follows.
First we must specify a one-parameter family of irreducible physical
acausal subspaces $\L(\tau)$.  On the basis of the argument in the
previous subsection we assume that $D(\L(\tau))$ is independent of
$\tau$. This corresponds to a family of acausal submanifolds in the
classical theory. In order to fix the reduction completely, we must
further specify a family of unitary mappings among the acausal
subspaces, which corresponds the mapping generated by the reduction
field in the classical theory. Without loss of generality we can
assume that this family of mappings is given by a family of unitary
operators $U(\tau)$, each of which maps $\L(0)$ to $\L(\tau)$.
Then the physical subalgebra $\B(\tau)$ for $\L(\tau)$ is given by
$\B(\tau)=U(\tau)\B(0)U(\tau)^{-1}$.

Now let $\Theta(\tau)$ be the causal mapping from $\L(0)$ to
$\L(\tau)$, which is conformal from \Theorem{theorem:conformality1}.
Further for $A\in\B(0)$ let $A(\tau)$ be the operator in $\B(0)$
defined by
$$
A(\tau):=\Theta(\tau)^{-1}U(\tau)AU(\tau)^{-1}\Theta(\tau).
$$
The for each probability amplitude $\Psi$ satisfying the weak
hamiltonian constraints and normalized on $\L(0)$,
$\Theta(\tau)^*\Psi|_{\L(\tau)}$ becomes independent of $\tau$ and
represented by a vector $\Phi$ in $\L(0)$. Further from
\Eq{Def:ExpectationValue:Pure1} the expectation
value of $A(\tau)$ with respect to this vector is expressed by
$$
(\Phi,A(\tau)\Phi)=<U(\tau)AU(\tau)^{-1}>_{\L(\tau)}.
$$
Hence, noting that $U(\tau)AU(\tau)^{-1}$ represents the operator at
the instant $\L(\tau)$ corresponding to the same measurement procedure
as for $A$ at $\L(0)$, the quantum dynamics in the totally constraint
system is mapped by reduction to the ordinary quantum dynamics in the
Hilbert space $\H_0:=\L(0)$ in the Heisenberg representation, where
$\Phi$ and $A(\tau)$ correspond to the invariant state and the
Heisenberg operator, respectively. Further since $U(\tau)^{-1}\Theta(\tau)$
is a conformal transformation on $\H_0$, there exists a family of
unitary operators $V(\tau)$ on $\H_0$, which is unique up to a phase
factor dependent on $\tau$, such that
$$
A(\tau)=V(\tau)^{-1}A V(\tau).
$$
Hence if we define the self-adjoint operator $h_0(\tau)$ by
$$
h_0(\tau)=iV(\tau)^{-1}{d V(\tau)\over d\tau},
$$
$A(\tau)$ follows the Heisenberg equation
$$
{d A(\tau)\over d\tau}=i[h_0(\tau),A(\tau)].
$$

Some comments are in order. Firstly the family $U(\tau)$ of unitary
operators is not unique even if we fix the acausal subspaces
$\L(\tau)$. Actually for any one-parameter family of self-adjoint
operators on $\H_0$ we can find an appropriate family $U(\tau)$ such
that $h_0(\tau)$ coincides with the given family of self-adjoint operators.
Hence we can pick up a special reduction only when $U(\tau)$
corresponds to some kind of symmetry of the system as in the case of
classical theory. We will give such examples in the next section.

Secondly for two different families of irreducible physical acausal
subspaces, $\L_1(\tau)$ and $\L_2(\tau)$, $D(\L_1(0))$ and $D(\L_2(0)$
do not coincide with each other in general. In such a case the causal
mapping between the reduced state spaces, $\L_1(0)$ and $\L_2(0)$,
yields a conformal mapping only between the proper closed subspaces,
$\L_1(0)\cap D(\L_2(0))$ and $\L_2(0)\cap D(\L_1(0))$, and it cannot
be extended to the whole reduced subspaces. This implies that reduced
quantum theories for the same classical totally constrained system are
not mutually unitary equivalent in the ordinary sense in general.

Finally there does not exist a self-adjoint operator $T$ such that the
family of irreducible physical acausal subspaces are given by the
eigenspaces of $T$, though they may be eigenspaces of a common
non-self-adjoint operator. For example, the family of acausal subspaces
\Beq
\L(\tau)=\SetDef{ \exp(-(x^0-\tau)^2/\sigma^2) \Phi(\bm{x})}
{\Phi(\bm{x})\in L_2(\RF^N)}
\Eeq
are irreducible acausal subspaces for the simple totally constrained
system with the single constraint operator $h=p_0$, and $\L(\tau)$ is
the eigenspace of the non-self-adjoint operator $T=x^0+i\sigma^2 p_0$.
Thus the instant operators $E(\tau)$ for $\L(\tau)$ do
not correspond to a time variable in the ordinary sense. This is not a
difficulty in practical situations, however, since it is sufficient if
we know which instant each state vector belongs. For example, the
reading of a clock is not a eigenvalue of some self-adjoint operator
but an index of the states of the clock.

\section{Example: Single Constraint Systems}

In this section we apply the abstract formalism developed in the
previous section to some simple totally constrained systems with a single
hamiltonian constraint in order to illustrate how the basic concepts such as
the central decomposition of the algebra of constants of motion, the physical
subalgebra, the irreducible physical acausal subspace and the causal mapping
are realized in concrete examples, and how the quantum dynamics is described
in terms of them.

Before going to the explicit examples, we first note that in our formalism
the quantum dynamics of two totally constrained systems have mathematically
the same structure if the $\VNA$-algebra $\W$ and its subalgebra $\I$ is
isomorphic, although their physical interpretations may be different.
In particular
for single constraint systems the spectrum and its multiplicity of
the constraint operator $h$ completely determines the structure of quantum
dynamics because two self-adjoint operators with the same spectrum and
multiplicity are mutually unitary equivalent\CITE{Conway.J1985B}.

\subsection{Central decomposition and the generalized eigenvectors of the
constraint operator}

In our formalism we must construct the central decomposition of $\C$ in order
to find the physical subalgebras and the corresponding irreducible physical
acausal subspaces. In the case of the single constraint systems this problem
is reduced to find a complete basis of the generalized eigenvectors of the
constraint operator as the following proposition shows.

\begin{proposition}\label{prop:CentralDecomposition:single}
Let $h$ be a self-adjoint operator with a dense domain on a Hilbert space $\H$,
and $\tilde\H:=\int_\sRF\oplus \H(\lambda)d\mu(\lambda)$ be a direct integral
Hilbert space with a finite regular Borel measure $\mu$ on $\RF$.
If $u_\lambda$ be a family of mappings from a linear
manifold $\D(\supseteq \domain{h}\cup\range{h})$ to $\H(\lambda)$
satisfying the conditions
\begin{itemize}
\item[a)] for each $x\in \D$, $x(\lambda):=u_\lambda(x)$ is defined
$\mu(\lambda)$-a.e. on $\RF$ and $||x(\lambda)||^2$ is integrable with
respect to $\mu$, i.e., $x(\lambda)$ defines a vector $Ux$ in $\tilde\H$,
\item[b)] $U:\D\maps\tilde\H$ is linear and $U\D$ is dense in $\tilde\H$,
\item[c)] $(x_1,x_2)=\int_\sRF d\mu(\lambda)(u_\lambda(x_1),u_\lambda(x_2))$
for any $x_1,x_2\in\D$,
\item[d)] for any $x\in\domain{h}$, $u_\lambda(hx-\lambda x)=0$ in $\lambda$
($\mu$-a.e.),
\end{itemize}
$U$ is uniquely extended to a unitary mapping from $\H$ to $\tilde\H$, and
gives a central decomposition of $\C=\VNA(h)'$:
$$
\Mapping{\C}{A}{\int_\sRF \oplus \C(\lambda)d\mu(\lambda)}
{\int_\sRF \oplus A(\lambda)d\mu(\lambda)},
$$
where $A(\lambda)u_\lambda(x)=u_\lambda(Ax)$ for $x\in\D$.
\end{proposition}
\begin{proof}
Clearly from conditions a)-c) $U$ is uniquely extended to a unitary
mapping from $\H$ to $\tilde \H$, and gives a direct integral representation
of $\H$. Let $E(\Delta)$ be the spectral measure of $h$.  Then for any
$x\in\H$ and any bounded Borel subset $\Delta$ of $\RF$, $E(\Delta)x$
is contained in $\domain{h}$. Further for any bounded Borel function
$f(\lambda)$ on $\RF$, there exists a sequence of polynomials $P_n(\lambda)$
such that
$$
\SOT-\lim_n E(\Delta)P_n(h)x=E(\Delta)f(h)x.
$$
Since $u_\lambda(P_n(h)E(\Delta)x)=P_n(\lambda)u_\lambda(E(\Delta)x)$
from condition d) and $U$ is unitary, from this it follows that
$$
\chi U(E(\Delta)f(h)x)=\chi f U(E(\Delta)x),
$$
where $\chi$ is the characteristic function of $\Delta$, and $\chi$ and $f$
are regarded as the diagonal bounded operators $\chi(\lambda)$ and $f(\lambda)$
in the direct integral representation. Hence taking the
limit $\Delta \tend \RF$, we obtain
$$
U(f(h)x)=f U(x).
$$
This means that $U$ maps the center of $\C$, $\VNA(h)$, onto the $\VNA$-algebra
of all the diagonal operators on $\tilde\H$.
\QED
\end{proof}

In the direct integral representation $\H(\lambda)$ is assumed to be
contained in a common Hilbert space $\hat\H$. Let $k$ be a vector in $\hat\H$.
Then if we regard $v_{\lambda,k}:=(k,u_\lambda)$ as the linear functional on
$\D$ defined by
\Beq
<v_{\lambda,k}, x>=(k, u_\lambda(x)),
\Eeq
condition d) of the proposition implies that $v_{\lambda,k}$ is a weak solution
to
\Beq
hv=\lambda v,
\Eeq
or a generalized eigenvector of $h$. Further condition c) represents the
completeness of the generalized eigenvectors $v_{\lambda,k}$. This viewpoint
is useful in the following examples.

\subsection{Non-relativistic quantum mechanics}

A classical canonical system with the phase space
$\RF^{2N}(\ni(\bm{x},\bm{p}))$ and a hamiltonian $h_0(\bm{x},\bm{p})$ is
embedded into a totally constrained system with a phase space
$\RF^{2N+2}(\ni(x^0,\bm{x},p_0,\bm{p}))$ and a single hamiltonian
constraint $h=p_0+h_0$\CITE{Kodama.H1995}.
Let us investigate the quantum dynamics of a
system $(\H=L_2(\RF^{N+1}), h)$ obtained by quantizing it by the
standard quantization procedure.

First of all note that by the unitary transformation
\Beq
U_0=\exp(ix^0 h_0),
\Eeq
the constraint operator $h$ is transformed as
\Beq
U_0 h U_0^{-1}=p_0.
\Eeq
Hence the structure of the relevant $\VNA$-algebras and their central
decomposition are determined by studying the case of vanishing hamiltonian.
Therefore we first consider the case $h=p_0$.

In the momentum representation the inner product of two state vectors
$\Phi_1(p_0,\bm{p})$ and $\Phi_2(p_0,\bm{p})$ are given by
\Beq
(\Phi_1,\Phi_2)=\int_\sRF d\lambda \int_{\sRF^N} \bar\Phi_1(\lambda,\bm{p})
\Phi_2(\lambda,\bm{p}).
\Eeq
Hence if we define $u_\lambda$ by
\Beq
u_\lambda: \Phi \maps \Phi(\lambda,\bm{p})\in L_2(\RF^N)=\H(\lambda),
\Eeq
it gives the central decomposition of $\H$ with respect to $\C$ from
\Prop{prop:CentralDecomposition:single}.  Hence for $h=p_0$ the $\VNA$-algebra
of constants of motion is given by
\Beq
\C=\VNA(p_0,\bm{x},\bm{p})
\Eeq
as is expected, and the physical subalgebras by
\Beq
\B=U\B_0 U^{-1}; \quad \B_0=\VNA(\bm{x},\bm{p}),
\Eeq
where $U$ is a unitary operator in $\C$.

{}From \Prop{proposition:StructureOfIPAS} irreducible acausal subspaces
invariant under $\B_0$ must be of the form
\Beq
\L_0(\chi)\cong\chi(x^0)\otimes L_2(\RF^N)
=\SetDef{\chi(x^0)\phi(\bm{x})}{\phi(\bm{x})\in L_2(\RF^N)}
\Eeq
in the coordinate representation where $\chi(x^0)\in L_2(\RF)$.
Since $\hat\N$ for $h=p_0$ is given by
\Beq
\hat\N=\SetDef{\xi'(x^0)}{\xi(x^0),\xi'(x^0)\in L_2(\RF)},
\Eeq
the acausality condition is expressed as
\Beq
\int_{-\infty}^\infty dx^0\,\chi(x^0)\not=0.
\Eeq
Hence $\L_0(\chi_1)$ and $\L_0(\chi_2)$ are causally related if and only if
there exists a constant $c$ such that
\Beq
\int_{-\infty}^\infty dx^0 (\chi_1(x^0)-c\chi_2(x^0))=0.
\label{CausalRelation:QM}\Eeq
In this case the causal mapping is bijective from $\L_0(\chi_1)$ onto
$\L_0(\chi_2)$ and expressed as
\Beq
\Theta(\chi_1\phi)=c \chi_2\phi.
\Eeq
Hence if $\chi_1$ and $\chi_2$ is normalized to unity in $L_2(\RF)$, $|c|$
yields the conformal factor of the mapping.

Here note that the instant operator for $\L_0(\chi)$ belongs to $\W(x^0,p_0)$,
but is not a function of $x^0$. Hence the value of $x^0$ should have some
uncertainty $\Delta x_0$ on each irreducible physical acausal subspace.
In practical situations this uncertainty is taken to be finite, which implies
that $\chi\in L_1(\RF)$. If we restrict $\L_(\chi)$ to such cases, they are
always causally related with each other from the above argument.

A general irreducible physical acausal subspace is given by the unitary
transformation in $\C$ from $\L_0(\chi)$ and is expressed in the direct
integral representation as
\Beq
\L_0(\chi,U)\cong \SetDef{\int_\sRF\oplus \hat\chi(\lambda)\hat
U(\lambda)\phi(\bm{x})}{\phi(\bm{x})\in L_2(\RF^N)},
\Eeq
where $\hat U(\lambda)$ is a measurable family of unitary operators in
$\B(L_2(\RF^N))$ and $\hat\chi(\lambda)$ is the Fourier transform of
$\chi(x^0)$. From the comment below \Cor{cor:SeparationOfNullSector}
the causal structure is determined only by the behavior of vectors around
$\lambda=0$ in the central decomposition. Hence if there exists a unitary
operator $\hat U(0)$ and a measurable family of bounded operators $A(\lambda)$
such that
\Beq
\hat U(\lambda)=\hat U(0) + \lambda A(\lambda),
\Eeq
we obtain the causal mapping
\Beq
\Theta:\Mapping{\L_0(\chi,U)}{U(\chi\otimes\phi)}
{\L_0(\chi)}{\chi\otimes \hat U(0)\phi},
\Eeq
which is bijective and unitary. On the other hand if the operator $A(\lambda)$
is not bounded around $\lambda=0$, the causal mapping is not defined on the
whole $\L_0(\chi,U)$ in general.  For example, let us consider
$U=\exp(ip_0 x^1)$. Since $U$ transforms $x^0$ as $U x^0 U^{-1}=x^0+x^1$,
$\L_0(\chi,U)$ now corresponds to the time variable $x^0+x^1$,
roughly speaking.
For this choice $\hat U(0)$ should be taken to be unity and $A(0)=i x^1$,
which is not bounded. Hence $\Theta$ is defined only in a dense region.
In this case, however, $\Theta$ can be uniquely extended to a unitary
mapping because it is isometric and defined in a dense region. This example
shows that if $d\hat U(\lambda)/d\lambda$ defines a densely defined operator,
$\Theta$ can be extended to a unitary mapping.

The description for $h=p_0$ so far is easily transferred to the non-trivial
hamiltonian case by the unitary transformation $U_0^{-1}$. In particular
the direct integral representation with respect to the central decomposition
coincides with that for the case $h=p_0$.  Hence the $\VNA$-algebra $\C$ is
given by
\Beq
\C=U_0^{-1}\VNA(p_0,\bm{x},\bm{p}) U_0,
\Eeq
the physical subalgebras by
\Beq
\B=U_0^{-1}U \VNA(\bm{x},\bm{p}) U^{-1}U_0,
\Eeq
and the irreducible physical acausal subspaces by
\Beq
\L(\chi,U)=U_0^{-1}\L_0(\chi,U)=U_0^{-1}U\L_0(\chi),
\Eeq
where $U$ is a unitary operator in $\VNA(p_0,\bm{x},\bm{p})$.

Even for the simplest choice of $U=1$ the instant operators belong to
$\VNA(x^0,h)$ and are not represented by a function of $x^0$ as in the
case of $h=p_0$.
Hence each irreducible physical acausal subspace
does not correspond to a subspace with constant $x^0$ unlike the naive
expectation. Nevertheless we can find a natural correspondence between
the present formalism with the usual formulation of quantum mechanics
in terms of the reduction discussed in the previous section.

To see this, first note that in the present case there is a natural family
of unitary transformations corresponding to the time translation. It is
given by
\Beq
U(\tau):=e^{-i\tau p_0},
\Eeq
which transforms $x^0$ as
\Beq
U(\tau)x^0 U(\tau)^{-1}=x^0-\tau.
\Eeq
Let $\L(\tau)$ be the family of irreducible physical acausal subspaces
given by
\Beq
\L(\tau):=U(\tau)\L(\chi,U)=U(\tau)U_0^{-1}U\L_0(\chi).
\Eeq
Then, since
\Beq
U(\tau)U_0^{-1}\Phi=U_0^{-1}e^{-i\tau(p_0-h_0)}\Phi\sim U_0^{-1}e^{i\tau
h_0}\Phi
\Eeq
for any vector $\Phi\in\H$, we find that the causal mapping $\Theta(\tau)$
from $\L(0)$ to $\L(\tau)$ is given by
\Beq
\Theta(\tau)(e^{i\tau h_0}\Phi)=U(\tau)\Phi.
\label{CausalMapping:QM}\Eeq
Hence the reduced dynamics on $\L(0)$ is represented by the unitary
transformation
\Beq
V(\tau)=U(\tau)^{-1}\Theta(\tau)=e^{-i\tau h_0},
\Eeq
and the corresponding Heisenberg operator $A(\tau)=V(\tau)^{-1}A V(\tau)$
on $\L(0)$ follows the Heisenberg equation
\Beq
{d A(\tau)\over d\tau}=i[h_0,A(\tau)].
\Eeq
Since $\L(0)=U_0^{-1}U\L_0(\chi)$ is isomorphic to
$L_2(\RF^N)=\{\phi(\bm{x})\}$, this equation can be rewritten as
the equation for $\hat A(\tau):=U^{-1}U_0A(\tau)U_0^{-1}U\in\B(L_2(\RF^N))$,
\Beq
{d \hat A(\tau)\over d\tau}=i[\hat h_0,\hat A(\tau)],
\Eeq
where $\hat h_0=U^{-1}h_0 U$. In particular for $U=1$ for which the instant
operator of $\L(\tau)$ belongs to $\VNA(x^0,h)$, this equation coincides
with the Heisenberg equation in the standard formulation of quantum mechanics.
On the other hand if we take $U=\exp(ip_0 x^1)$, we obtain quantum mechanics
with respect to the time variable $x^0+U_0^{-1}x^1U_0$. These two gauges are
unitary equivalent as noted above. However, for example, if we take
$U=\exp(ix^1\sin(1/p_0))$, which is a well-defined unitary operator
corresponding to the time variable $x^0-(U_0^{-1}x^1U_0/h^2)\sin(1/h)$,
we obtain a  non-unitary-equivalent theory.

Finally note that for the family of acausal subspaces $\L(\tau)$,
from \Eq{CausalMapping:QM} the relative probability amplitude $\Psi$
satisfies
\Beq
\Psi(U(\tau)\Phi)=\Psi(\Theta(\tau)e^{i\tau h_0}\Phi)=\Psi(e^{i\tau h_0}\Phi)
\Eeq
for any $\Phi\in\L(0)$. Hence if we define the  vector $\Phi(\tau)$
in $\L(0)$ by
\Beq
<\Phi(\tau),\Phi>=\Psi(U(\tau)\Phi) \quad \forall \Phi\in\L(0),
\Eeq
it is related to $\Phi(0)$ by
\Beq
\Phi(t)=e^{-i\tau h_0} \Phi(0).
\Eeq
This implies that $\Phi(t)$ satisfies the standard Schr\"{o}dinger equation.

\subsection{Relativistic free particle}

Classical dynamics of a relativistic free particle in Minkowski spacetime
is described by a totally constrained system on the phase space
$\Gamma=\{(x^\mu,p_\mu)\in \RF^4\}$ with constraint
\Beq
h={1\over2}(p_\mu p^\mu + m^2),
\Eeq
where $m$ is the mass\CITE{Kuchar.K1981B,Kodama.H1995}. The quantum theory for
this
system in our formulation is given as follows.

The inner product of state vectors in $\H=L_2(\RF^4)$ is written as
\Beqr
&(\Phi_1,\Phi_2)&=\int d^4p \overline{\hat\Phi_1(p)}\hat\Phi_2(p)
=\int_{-\infty}^\infty d\lambda\int d^4p \, \delta(p^2+\lambda)
\overline{\hat\Phi_1(p)}\hat\Phi_2(p) \nonumber\\
&&=\int_0^\infty d\lambda \int d^3p {1\over2\omega_\lambda}
[\overline{\hat\Phi_1(\omega_\lambda,\bm{p})}\hat\Phi_2(\omega_\lambda,\bm{p})
+\overline{\hat\Phi_1(-\omega_\lambda,\bm{p})}\hat\Phi_2(-\omega_\lambda,\bm{p}) \nonumber\\
&&\quad +\int^0{-\infty} d\lambda \int_{-\infty}^\infty dp_0\int_{S^2}d\Omega
{|\bm{p}|_\lambda\over2}\overline{\hat\Phi_1(p_0,|\bm{p}|_\lambda\Omega)}
\hat\Phi_2(p_0,|\bm{p}|_\lambda\Omega),
\Eeqr
where $\hat\Phi_j(p)$ is the momentum representation of $\Phi_j$ and
\Beq
\omega_\lambda:=\sqrt{|\bm{p}|^2+\lambda}, \quad
|\bm{p}|_\lambda:=\sqrt{p_0^2-\lambda}.
\Eeq
Hence if we define the mapping $u_\lambda$ by
\Beq
\begin{array}{ll}
u_\lambda:\Mapping{\H}{\Phi}{\H_T=L_2(\RF^3)\oplus L_2(\RF^3)}
{\left({1\over\sqrt{2\omega_\lambda}}\hat\Phi(\omega_\lambda,\bm{p}),
{1\over\sqrt{2\omega_\lambda}}\hat\Phi(-\omega_\lambda,\bm{p})\right)},
& \r{for}\quad \lambda>0, \\
u_\lambda:\Mapping{\H}{\Phi}{\H_S=L_2(\RF\times S^2)}
{\sqrt{|\bm{p}|_\lambda/2}\hat\Phi(p_0,|\bm{p}|_\lambda\Omega)}
& \r{for}\quad \lambda<0,
\end{array}
\Eeq
$u_\lambda$ satisfies the conditions of \Prop{prop:CentralDecomposition:single}
for the self-adjoint operator $\Lambda$ defined by
\Beq
\Lambda=-p_\mu p^\mu\equiv -p^2.
\Eeq
with $d\mu(\lambda)=d\lambda$ and $\H(\lambda)=\H_T(\lambda>0),
\H_S(\lambda<0)$. Therefor $U\Phi=\int\oplus u_\lambda(\Phi)d\lambda$
yields the direct
integral representation of $\H$ corresponding to the central decomposition
of $\C$, and $h$ is expressed as $(m^2-\lambda)/2$ in this representation.

As is clear from this construction, $\H$ is written as a direct sum
\Beq
\H\cong L_2(\RF_+,\H_T) \oplus L_2(\RF_-,\H_S),
\Eeq
which induces the  decomposition
\Beq
\C = \C_T \oplus \C_S,
\Eeq
where
\Beqr
&&\C_T\cong L_\infty(\RF_+)\bar\otimes \B(\H_T),\\
&&\C_S\cong L_\infty(\RF_-)\bar\otimes \B(\H_S).
\Eeqr

$\B(\H_T)$ is generated by the self-adjoint operators
\Beq
\bm{p},\quad -i\nabla_{\bm{p}}, \quad
\Pi_+:=\begin{matrix}{cc}1&0\\0&0\end{matrix}, \quad
T:=\begin{matrix}{cc}0&1\\1&0\end{matrix}.
\Eeq
It is easily checked using the definition of $u_\lambda$ that these correspond
to the product of the projection operator $E_T:=\theta(-p^2)$ and
the self-adjoint operators on $\H$ given by
\Beq
\bm{P}:=\bm{p}, \quad \bm{X}:=\bm{x}+{\bm{p}\over p_0}(x^0+{i\over2p_0}),
\quad \Pi_+, \quad T,
\Eeq
where $\Pi_+$ and $T$ represent the projection operator onto the standard
positive frequency part and the time reversal unitary operator on $\H$,
respectively.
Hence the restriction of a physical subalgebra $\B$ on the $T$-section
$L_2(\RF_+,\H_T)$ is given by
\Beq
E_T \B:=U_T E_T\VNA(\bm{P},\bm{X},\Pi_+,T)U_T^{-1},
\Eeq
where $U_T$ is a unitary operator in $E_T\C=\C_T$.

On the other hand $\B(\H_S)$ is generated by the  self-adjoint operators
\Beq
p^0,\quad -i\partial_{p_0},\quad \bg{\Omega}, \quad
i\bg{\Omega}\times\nabla_{\bg{\Omega}},
\Eeq
which correspond to the product of the projection operator
$E_S:=\theta(p^2)$ and the self-adjoint operator on $\H$ given by
\Beq
P_0:=p_0,\quad
X^0:=x^0+{p_0\over|\bm{p}|^2}\left(\bm{p}\cdot\bm{x}-{i\over2}\right), \quad
{\bm{p}\over|\bm{p}|}, \quad
\bm{J}:=-i\bm{p}\times\bm{x}.
\Eeq
Hence the restriction of the physical subalgebra $\B$ on the $S$-sector
$L_2(\RF_-,\H_S)$ is given by
\Beq
E_S\B=U_S E_S\VNA(P_0,X^0,\bm{p}/|\bm{p}|,\bm{J}) U_S^{-1},
\Eeq
where $U_S$ is a unitary operator in $E_S\C=\C_S$.

Since any two separable Hilbert spaces of infinite dimension are
isomorphic, $\H_T$ and $\H_S$ are isomorphic.  Let $f$ be such an isomorphism.
Then $f$ induces a $\VNA$-isomorphism $f_*$
\Beq
f_*: E_T\VNA(\bm{P},\bm{X},\Pi_+,T)\maps
E_S\VNA(P_0,X^0,\bm{p}/|\bm{p}|,\bm{J}).
\Eeq
The physical algebra $\B$ is expressed in terms of this isomorphism as
\Beq
\B=\SetDef{ U_TAU_T^{-1}+ U_Sf_*(A) U_S^{-1}}
{A\in E_T \VNA(\bm{P},\bm{X},\Pi_+,T)}.
\Eeq
We will denote $\B$ corresponding to $U_T=E_T$ and $U_S=E_S$ for some $f$
by $\B_0$ in this section. The generic $\B$ above is written in terms of
$\B_0$ as
\Beq
\B=U\B_0 U^{-1},
\Eeq
where $U:=U_T\oplus U_S$ represents an arbitrary unitary transformation in
$\C$.

{}From \Prop{proposition:StructureOfIPAS} an irreducible physical acausal
subspace for $\B_0$ must have the form
\Beq
\L(\chi)\cong \SetDef{\chi_+\otimes \phi \oplus
\chi_-\otimes f(\phi)}
{\phi=(\phi_+,\phi_-)\in\H_T},
\Eeq
where $\chi_\pm\in L_2(\RF_\pm)$. Since the constraint operator $h$
is expressed by the function $(m^2-\lambda)/2$ in the central decomposition,
the acausality condition is expressed as
\Beq
\int_{-\infty}^\infty d\lambda {|\chi(\lambda)|^2\over
(\lambda-m^2)^2}=+\infty,
\Eeq
where $\chi=\chi_++\chi_-$. This shows that the acausality condition
only restrict $\chi_+$ for $m^2>0$. This reflects the fact that
$h$ is invertible in the $S$-sector, hence $E_S\H\subset\N$.

The generic irreducible physical acausal subspace for the physical
subalgebra $\B=U\B_0 U^{-1}$ is given by
\Beq
\L(\chi,U):=U\L(\chi).
\Eeq
In the momentum representation a vector $\Phi$ in $\L(\chi,U)$ is
expressed in terms of $\phi\in\H_T$ as
\Beqr
&&\hat\Phi(p)=\chi(-p^2)\left[\theta(-p^2)\sqrt{2|p_0|}\tilde\phi_T(p)
+\theta(p^2)\sqrt{2\over|\bm{p}|}\tilde\phi_S(p)\right];
\label{RFP:Phi(p)}\\
&&\tilde\phi_T(p)=\theta(p^0)(U_T(-p^2)\phi)_+(\bm{p})
+\theta(-p^0)(U_T(-p^2)\phi)_+(\bm{p}),\\
&&\tilde\phi_S(p)=(U_S(-p^2)f(\phi))(p_0,{\bm{p}\over|\bm{p}|}).
\Eeqr

As in the nonrelativistic particle case, the causal relation among
the irreducible acausal subspaces are determined by the behavior of
$\chi(\lambda)$ and $U(\lambda)$ around $h(\lambda)=0$. In particular
the causal mapping between two acausal subspaces $\L(\chi_1)$ and $\L(\chi_2)$
are bijective and conformal if and only if
\Beq
\int_0^\infty{|\chi_{1+}(\lambda)-\chi_{2+}(\lambda)|^2\over(\lambda-m^2)^2}
d\lambda < +\infty.
\Eeq
Further if $U_T(\lambda)$ is written around $\lambda=m^2$ in terms of a
family of operators $A(\lambda)$ with a dense domain as
\Beq
U_T(\lambda)-U_T(m^2)=(\lambda-m^2)A(\lambda),
\Eeq
the causal mapping from $\L(\chi,\U)$ to $\L(\chi)$ can be extended to
a unitary mapping.

For example, for the unitary operator $U(a)=e^{-ip\cdot a}$ which represents
the spacetime translation,
\Beq
U(a)x^\mu U(a)^{-1}=x^\mu -a^\mu,
\Eeq
we obtain
\Beq
U_T(a)(\chi_+(\lambda)\phi)\sim \chi_+(\lambda)V(a)^{-1}\phi,
\Eeq
where
\Beq
V(a)=\begin{matrix}{cc}e^{-i\omega a^0+i\bm{p}\cdot\bm{a}}&0\\
0&e^{i\omega a^0+i\bm{p}\cdot\bm{a}}\end{matrix}.
\Eeq
Here $\omega=\sqrt{\bm{p}^2+m^2}$.
Hence if we define the mapping $F:\H_T\maps\L(\chi)$ by
\Beq
F(\phi)\cong \chi_+\otimes \phi\oplus \chi_-\otimes f(\phi),
\Eeq
the causal mapping $\Theta:\L(\chi)\maps U(a)\L(\chi)$ is represented as
\Beq
F^{-1}U(a)^{-1}\Theta F(\phi)=V(a)\phi.
\Eeq
In particular for a one-parameter family of irreducible physical acausal
subspaces $\L_\tau:=U(a(\tau))\L(\chi)$, the quantum dynamics is reduced
to the Schr\"{o}dinger equation for the state vector
$\phi(\tau)=V(a(\tau))\phi$ in $\H_T$ given by
\Beq
i{d\over d\tau}\phi(\tau)=\begin{matrix}{cc}
\dot a_0\omega-\dot {\bm{a}}\cdot\bm{p}&0\\
0 & -\dot a_0\omega-\dot{\bm{a}}\cdot\bm{p}\end{matrix} \phi(\tau).
\Eeq

In a similar way we can find the causal relation between two irreducible
physical acausal subspaces related by a Lorentz transformation.
Let $M_{\mu\nu}$ be the generator of Lorentz transformations defined by
\Beq
M_{\mu\nu}=x_\mu p_\nu - x_\nu p_\mu.
\Eeq
Then for an arbitrary constant antisymmetric tensor $b^{\mu\nu}$
we obtain
\Beq
{1\over2}M_{\mu\nu}b^{\mu\nu}=-p_0\left(\bm{X}+{i\over
2p_0^2}\bm{P}\right)\cdot\bm{b}_B
+(\bm{X}\times\bm{P})\cdot \bm{b}_R,
\Eeq
where $b_B^j=b^{0j}$ and $b_R^j=\epsilon^{jkl}b_{kl}$.  Hence the
transformation induced on $\H_T$ by
\Beq
U(b):=\exp\left(-i{1\over2}M_{\mu\nu}b^{\mu\nu}\right)
\Eeq
and the causal mapping $\Theta:\L(\chi)\maps U(b)\L(\chi)$ is given by
the unitary operator
\Beq
V(b):=F^{-1}U(b)^{-1}\Theta F
=\begin{matrix}{cc} e^{iA_+}&0\\ 0 & e^{iA_-}\end{matrix},
\Eeq
where $A_\pm$ are the self-adjoint operators defined by
\Beq
A_\pm:=\pm\omega\left(\bm{X}+{i\over2\omega^2}\bm{P}\right)\cdot\bm{b}_B
+(\bm{X}\times\bm{P})\cdot\bm{b}_R.
\Eeq

The weak hamiltonian constraint on the relative probability amplitude $\Psi$
is a dual of the conventional Dirac constraint. Hence when $\Psi(\Phi)$
can be written as
\Beq
\Psi(\Phi)=\int d^4x\,\bar\Psi(x)\Phi(x),
\label{RFP:Psi(x):def}\Eeq
$\Psi(x)$ is a generalized solution to the Klein-Gordon equation
\Beq
(-\partial_\mu\partial^\mu + m^2)\Psi(x)=0.
\Eeq
Hence the norm $||\Psi||_\L$ on an irreducible physical acausal subspace
introduces a positive definite norm to the linear space of solutions to
the Klein-Gordon equations. Let us investigate the relation between this
norm and the Klein-Gordon product defined by
\Beq
N(\Psi_1,\Psi_2):=i\int d^3x\,\bar\Psi_1(\bm{x},x^0)
\mathop{\partial_0}\limits^\leftrightarrow\Psi_2(\bm{x},x^0).
\Eeq

Let $\phi=(\phi_+(\bm{p}),\phi_-(\bm{p}))$ be a vector $\H_T$, and express
$\Psi(F(\phi))$ as
\Beq
\Psi(F(\phi))=\int d^3p \left(\bar\Psi_+(\bm{p})\phi_+(\bm{p})
+\bar\Psi_-(\bm{p})\phi_-(\bm{p})\right).
\Eeq
Then the norm of $\Psi$ on the acausal subspace $\L(\chi)$ is expressed as
\Beq
||\Psi||^2_{\L(\chi)}= \int d^3p \left(|\Psi_+(\bm{p})|^2
+|\Psi_-(\bm{p})|^2\right).
\Eeq
On the other hand $\Psi(F(\phi))$ is rewritten as
\Beqr
&\Psi(F(\phi))&=
\int d^4\, \delta(p^2+m^2)2\omega\left(
\theta(p^0)\bar\Psi_+(\bm{p})\phi_+(\bm{p})
+\theta(-p^0)\bar\Psi_-(\bm{p})\phi_-(\bm{p})\right)\nonumber\\
&&={1\over\chi(m^2)}\int d^4\, \delta(p^2+m^2)\sqrt{2\omega}\left(
\theta(p^0)\bar\Psi_+(\bm{p})+\theta(-p^0)\bar\Psi_-(\bm{p})\right)\nonumber\\
&&\quad \times\chi(-p^2)\sqrt{2\omega}
\left(\theta(p^0)\phi_+(\bm{p})+\theta(-p^0)\phi_-(\bm{p})\right).
\Eeqr
Comparing this with \Eq{RFP:Psi(x):def} and \Eq{RFP:Phi(p)}
we find that $\Psi(x)$ is expressed in terms of
$\Psi_\pm(\bm{p})$ as
\Beq
\Psi(x)={1\over(2\pi)^2\chi(m^2)}
\int d^3p\,{1\over\sqrt{2\omega}}e^{i\bm{p}\cdot\bm{x}}
\left(e^{-i\omega x^0}\Psi_+(\bm{p}) + e^{i\omega x^0}\Psi_-(\bm{p})\right).
\Eeq
Hence we obtain
\Beq
||\Psi||^2_{\L(\chi)}=2\pi|\chi(m^2)|^2
\left(N(\Psi_+,\Psi_+)-N(\Psi_-,\Psi_-)\right),
\label{RFP:NormByKGproduct}\Eeq
where $\Psi_+(x)$ and $\Psi_-(x)$ are the positive and negative
frequency parts corresponding to $\Psi_+(\bm{p})$ and $\Psi_-(\bm{p})$,
respectively.

Since the causal mapping between $\L(\chi)$ and $\L(\chi,U)$ is unitary
as far as $U(\lambda)$ is regular around $\lambda=m^2$,
\Eq{RFP:NormByKGproduct} still holds for the norm defined on
$\L(\chi,U)$. This implies that our normalization of the relative probability
amplitude induced from the inner product of the Hilbert space automatically
picks up a decomposition of solutions to the Klein-Gordon equation
into special positive and negative frequency parts, which corresponds
to the usual Minkowski vacuum in the field theory.

This phenomenon is not specific to the Minkowski background, but occurs
in curved background in general if it is
Lorentzian(cf. Ref.\cite{Kuchar.K1991B}).
It is because
the constraint operator of such a system has a continuous spectrum
with infinite multiplicity, hence the system is unitary equivalent to
the present system as noted at the beginning of this section.

For example, let us consider a free relativistic particle moving
in a background spacetime
\Beq
ds^2=-dt^2 + a(t)^2 d\bm{x}^2.
\Eeq
The constraint operator $h$ of this system is given by
\Beq
2h=-(p_0)^2+a^{-2}\bm{p}^2 + m^2.
\Eeq

To find the central decomposition of this system, we need to solve the
generalized eigenvalue problem
\Beq
(\partial_t^2+ a^{-2}\bm{p}^2)u=-\lambda u.
\Eeq
Let $v^{(j)}(t,\bm{p})$($j=\pm$) be two independent solutions to this equation
and define the mapping $u_\lambda:\H \maps L_2(\RF^3)\oplus L_2(\RF^3)$ by
\Beqr
&&u_\lambda(\Phi)=\left(\phi_+(\lambda,\bm{p}),\phi_-(\lambda,\bm{p})\right),\\
&&\phi_\pm=\int_{-\infty}^\infty dt \overline{v^{(\pm)}_\lambda(t,\bm{p})}
\hat\Phi(t,\bm{p}).
\Eeqr
Then from \Prop{prop:CentralDecomposition:single} $u_\lambda$ defines
a central decomposition if and only if $v^{(\pm)}_\lambda$ satisfies
the following completeness relation
\Beq
\sum_{j=\pm}\int_{\Lambda(\bm{p})}d\lambda
v^{(j)}_\lambda(t_1,\bm{p}) \overline{v^{(j)}_\lambda(t_2,\bm{p})}
=\delta(t_1-t_2),
\Eeq
where $\Lambda(\bm{p})$ is the range of the spectrum for given $\bm{p}$.
In particular from this it follows the orthonormality condition
\Beq
\int_{-\infty}^\infty dt
\overline{v^{(j)}_\lambda(t,\bm{p})}v^{(j')}_{\lambda'}(t,\bm{p})
=\delta^{jj'}\delta(\lambda-\lambda').
\Eeq
If we require that
\Beq
\overline{v^{(+)}_\lambda(t,\bm{p})}=v^{(-)}_\lambda(t,-\bm{p}),
\Eeq
this orthonormality condition uniquely fixes $u^{(\pm)}_\lambda$
up to phase factors for each.  Hence the decomposition into the
positive and negative frequency parts is uniquely determined.

Of course the each modes determined by this are mixture of the natural
positive and negative frequency modes in the region where $a(t)$ is
approximately constant if the spacetime is not static. For example,
for
\Beq
a^2=[1+\alpha (1-\tanh^2 t)]^{-1}\quad (\alpha>0),
\Eeq
$v^{(+)}_\lambda$ is given by
\Beqr
&&v^{(+)}_\lambda(t,\bm{p})=\sqrt{\pi\omega\over 4\sinh\pi\omega}
\left(A(\nu,\omega)P^{i\omega}_\nu(\tanh t)
+B(\nu,\omega)P^{-i\omega}_\nu(\tanh t)\right);\nonumber\\
&&\\
&&A(\nu,\omega)=\left(1+{1\over\sqrt{1+\xi^2}}\right)^{1/2},\\
&&B(\nu,\omega)=-\left(1+{1\over\sqrt{1+\xi^2}}\right)^{-1/2}
{\sin\nu\pi\over \sin(\nu-i\omega)\pi}{\Gamma(1+\nu+i\omega)\over
\Gamma(1+\nu-i\omega)},
\Eeqr
where
\Beqr
&& \omega=\sqrt{\bm{p}^2+\lambda},\\
&& \nu(\nu+1)=\alpha\bm{p}^2,\\
&&\xi={\sin\nu\pi\over\sinh\pi\omega}.
\Eeqr
In the asymptotic region $t\tend \infty$
$(4\sinh\pi\omega/\pi\omega^{1/2}v^{(+)}_\lambda$ approaches
\Beq
A(\nu,\omega){e^{i\omega t}\over\Gamma(1-i\omega)}
+B(\nu,\omega){e^{-i\omega t}\over\Gamma(1+i\omega)},
\Eeq
and in the region $t\tend-\infty$,
\Beq
-{i\over\sqrt{1+\xi^2}}{\sin(\nu+i\omega)\pi\over\sinh\pi\omega}
{\Gamma(1+\nu+i\omega)\over\Gamma(1+\nu-i\omega)}
\left(A(\nu,\omega){e^{i\omega t}\over\Gamma(1+i\omega)}
+\overline{B(\nu,\omega)}{e^{-i\omega t}\over\Gamma(1-i\omega)}\right).
\Eeq

\subsection{Minisuperspace models}

The so-called minisuperspace models of general relativity also give
totally constrained system with a single constraint. The classical
hamiltonian constraints of minisuperspace models are all written in terms
of canonical pairs $(x^\mu,p_\mu)$ of reduced metric variables
and matter variables in the form
\Beq
h={1\over2}g^{\mu\nu}(x)p_\mu p_\nu + \V(x),
\Eeq
where the supermetric $g_{\mu\nu}(x)$ has the signature $[-,+,+,\cdots]$
in general\CITE{Kuchar.K1991B}. Hence they are equivalent to the system of a
relativistic
particle moving in a curved spacetime under the influence of the potential
$V(x)$. In particular its quantum dynamics in our formalism has mathematically
the same structure as that of a relativistic free particle discussed in
the previous section since the quantized constraint operator $h$ has
a continuous spectrum with infinite multiplicity except for the systems
with one degree of freedom. Therefore in this section we only discuss
problems which have not been touched upon so far.

In our formalism we defined the quantum totally constrained system as a
mathematically well-defined pair of a Hilbert space $\H$ and
a set of self-adjoint operators $\{h_\alpha\}$, and have not
discussed the procedure to get such a pair from a classical totally
constrained system. As is well-known, such quantization procedures are
not unique and one can get an infinite number of (potentially) different
quantum systems from the same classical system depending on the choice of
fundamental variables and operator orderings. This non-uniqueness is
unavoidable and cannot be removed only by theoretical considerations
in general. Here we point out another source of non-uniqueness related
to the assumption of self-adjointness on the constraint operators.

When a real function of classical canonical variables is given, it is usually
not so hard to find a corresponding operator which is symmetric
in a dense domain by an appropriate choice of an operator ordering
and a representation.  However, it is often difficult to find its
self-adjoint extensions and such extensions are not unique in general.
In such cases even if the representation and the operator ordering
of the constraints are fixed, different self-adjoint extensions of them
give different quantum dynamics in our formalism.

For example, let us consider the minisuperspace model of a spatially
homogeneous universe with a positive cosmological constant. If we take
the logarithm of the cosmic scale factor, $\alpha$, and its conjugate
momentum $p$ as the fundamental canonical variables, the constraint
function $h$ is given by
\Beq
h=f(\alpha,p)[-p^2 -K e^{-4\alpha}+\Lambda e^{6\alpha}],
\Eeq
where $K$ is the spatial curvature, $\Lambda$ is the cosmological
constant, and $f(\alpha,p)$ is an arbitrary positive
function\cite{Kodama.H1993}. Here for the technical simplicity we take
$f=e^{-4\alpha}$ and fix the operator ordering of the form
\Beq
h=e^{-\alpha}pe^{-2\alpha}pe^{-\alpha}+\Lambda e^{2\alpha}-K.
\Eeq
Then in the coordinate representation
\Beqr
&&\Phi=\Phi(\alpha)\in\H=L_2(\RF^2),\\
&&p=-i{d\over d\alpha},
\Eeqr
$h$ is represented as
\Beq
h\Phi={d\over d\alpha}\left(e^{-4\alpha}{d\Phi\over d\alpha}\right)
+(\Lambda e^{2\alpha}-K+3e^{-4\alpha})\Phi,
\Eeq
which is symmetric on the set of smooth rapidly decreasing functions.
When we introduce the variables
\Beqr
&& z={1\over2}\Lambda^{1/3}e^{2\alpha},\\
&& w(z)=e^{-\alpha}\Phi(\alpha),
\Eeqr
$h$ is transformed to
\Beq
(\Phi_1,h\Phi_2)=(2\Lambda)^{1/3}\int_0^\infty dz \overline{w_1(z)}
\left({d^2\over dz^2}+z -{K\over (2\Lambda)^{2/3}}\right)w_2(z).
\Eeq
Since the inner product is written as
\Beq
(\Phi_1,\Phi_2)=(2\Lambda)^{1/3}\int_0^\infty dz \overline{w_1(z)} w_2(z),
\Eeq
we easily see that $h$ is self-adjoint if and only if $w(z)$ satisfies
the condition that $\mu:=w(0)'/w(0)$ is a real constant.
Hence we obtain an infinite number of different self-adjoint extension
of $h$ depending on the choice of $\mu$($-\infty\le\mu\le\infty$).

This freedom in the self-adjoint extension affects the quantum dynamics.
In the present case for the boundary condition $w'(0)=\mu w(0)$ the
orthonormal set of weak solutions to
\Beq
h v_\lambda(\alpha)=\lambda v_\lambda(\alpha)
\Eeq
is given by
\Beq
v_\lambda(\alpha)={(2\Lambda)^{-1/3}\over\sqrt{1+\nu(\tilde\lambda)^2}}
e^\alpha\left[Ai(\tilde\lambda-(\Lambda/4)^{1/3}e^{2\alpha})
+\nu(\tilde\lambda)Bi(\tilde\lambda-(\Lambda/4)^{1/3}e^{2\alpha})\right],
\Eeq
where $Ai(x)$ and $Bi(x)$ are Airy functions, and
\Beqr
&&\tilde \lambda:=(2\Lambda)^{-2/3}(\lambda+K),\\
&&\nu(\tilde\lambda):=-{\mu Ai(\tilde\lambda)+Ai'(\tilde\lambda)
\over \mu Bi(\tilde\lambda)+Bi'(\tilde\lambda)}.
\Eeqr
Since the direct integral representation by the central decomposition
is given by
\Beqr
&&u_\lambda(\Phi):=\phi(\lambda)=\int_{-\infty}^\infty d\alpha
v_\lambda(\alpha)\Phi(\alpha),\\
&& \Phi(\alpha)=\int_{-\infty}^\infty d\lambda \phi(\lambda)v_\lambda(\alpha),
\Eeqr
from \Prop{prop:CentralDecomposition:single}, the acausal subspaces
and the linear null manifold depends on the choice of $\mu$.
The same problem occurs in systems with scalar fields.

Another subtlety of our formalism is the fact that the causal mapping
is conformal but not isometric in general. Let us consider the simple
example above again.  In this example the physical subalgebra is
trivial and the irreducible physical acausal subspaces are all
one-dimensional. Hence the quantum dynamics should be trivial in our
formalism, but the conformal factor behaves non-trivially.

Let $\L(\chi)$ be an irreducible physical acausal
subspace. Then in the direct integral representation $\Phi\in\L(\chi)$
is expressed as $\phi(\lambda)=c\chi(\lambda)$ where $c$ is a complex
constant and $\chi(\lambda)$ is a function such that $||\chi||=1$
and
\Beq
\int_{-\infty}^\infty d\lambda {|\chi(\lambda)|^2\over\lambda^2}=\infty.
\Eeq
For simplicity we assume that $\chi(\lambda)$ is continuous at $\lambda=0$.
Then this acausality condition is simply given by the condition
$\chi(0)\not=0$. If we express the relative probability amplitude $\Psi$
as
\Beq
\Psi(\Phi)=\int_{-\infty}^\infty d\alpha \overline{\Psi(\alpha)}\Phi(\alpha),
\Eeq
by the same argument as in the previous subsection we obtain
\Beq
\Psi(\alpha)=\chi(0)^{-1}\psi v_0(\alpha),
\Eeq
where $\psi$ is a complex number parametrizing $\Psi$ and is related to
the norm of $\Psi$ as
\Beq
||\Psi||_{\L(\chi)}=|\psi|.
\Eeq
Hence for two acausal subspaces $\L(\chi_1)$ and $\L(\chi_2)$ which are
conformally related, the norms of $\Psi$ on each subspaces are related by
\Beq
||\Psi||_{\L(\chi_2)}=|{\chi_2(0)\over\chi_1(0)}| ||\Psi||_{\L(\chi_1)}.
\Eeq
On the other hand for the causal mapping
\Beq
\Theta:\L(\chi_1) \maps \L(\chi_2),
\Eeq
its conformal factor is given by
\Beq
{||\Theta(\Phi)||\over ||\Phi||}=|{\chi_1(0)\over\chi_2(0)}|.
\Eeq
Hence the ratio of the norms of $\Psi$ coincides with the inverse
of the conformal factor.

Here note that due to the normalization $||\chi||=1$ the value $|\chi(0)|$
represents how well the hamiltonian constraint is satisfied for states
in the acausal subspaces $\L(\chi)$. Hence for acausal subspaces on which
the hamiltonian constraint is poorly satisfied the norm of $\Psi$
becomes small and the conformal factor becomes large. Thus the conformal
factor of the causal mapping seems to have some physical significance.
This point is neglected in our formalism because we have restricted
the comparison of the probability within state vectors in the same
irreducible physical acausal subspace.  Hence in our formalism it is
equally allowed to consider an acausal subspace for which
$\chi(\lambda)$ is concentrated around a value of $\lambda$ such that
$h(\lambda)\not=0$ if $\chi(0)\not=0$, i.e., the hamiltonian
constraint is quite poorly satisfied. This aspect may appear
unsatisfactory to some people, but it seems to be an inevitable
feature of the canonical approach to the author because in the
canonical approach the choice of an instant is not an object of
dynamics.

\section{Discussions}

In this paper we have developed a general framework to formulate the
quantum dynamics of totally constrained systems without gauge fixing or
without referring to special time variables by considering a unbounded
relative probability amplitude $\Psi$ and postulating the constraints as
the weak hamiltonian constraints on $\Psi$. In particular we have shown
that the quantum dynamics as regarded as causal mappings among irreducible
physical acausal subspaces is unitary or conformal if the $\VNA$-algebra
of constants of motion is of type I, which includes the abelian constrained
systems.

The weak hamiltonian constraints, which are the basic dynamical
equation in our formalism, are in a sense a dual form to the
conventional Dirac constraints, and at a first glance appear to be the
same as the latter.  In fact, in many arguments in which mathematical
rigor is neglected, they are not properly distinguished. However,
these two formulations are in reality quite different because in the
latter the states satisfying only the constraints are regarded as
physical and only the operators commuting with the constraints are
includes as observables, while in our formulation the constraints do
not restrict states nor observables by themselves.  This difference is
crucial in incorporating operators corresponding to time variables in
the classical theory as observables, and in making possible to discuss
dynamics in the quantum framework without losing link to measurements.

The most prominent feature of our formulation is its flexibility in
the description of dynamics. In particular, we can consider the gauge
fixing or the reduction and investigate the relation between two
reductions corresponding to different gauge fixings completely within
the quantum framework unlike in the usual approaches.  Further, though
we have investigated in detail the dynamics among irreducible physical
acausal subspaces in order to make sure that the formulation is
physically sensible, the framework itself allows us to discuss the
dynamics among incomplete acausal subspaces which are selected by
gauge conditions depending on states or relative probability
amplitude functionals.

Of course we had to pay some cost to allow this flexibility. In particular
in our formalism we had to give up determining the relative probability
of all the states in the state space
from a state vector given by measurements. Though this restriction
of prediction is partly caused by the implicit smoothness requirement of
dynamics as explained in \S3.6, it is also caused by the fact that
the instant operators do not commute with each other in general as we cannot
predict the probability distribution for the position measurement and
the momentum measurement at the same time simultaneously in quantum
mechanics.  Hence this restriction is inevitable as far as we allow the
state vectors carry information on instants.

Finally we would like to comment on the application of our formalism to
quantum gravity.  In principle our formalism, or more precisely, our
framework is applicable to quantum gravity regardless whether it is based
on general relativity or superstring theories.  However, in order to
investigate concrete dynamical problems of quantum gravity in our framework,
we must solve lots of practical problems in advance.  First of all,
we must explicitly construct the Hilbert space and the mathematically
well-defined representation of constrained operators on it because
they are the starting point of our framework.  For that purpose, for
example in the case of quantum gravity based on general relativity,
a mathematically well-defined classical canonical formalism must be
constructed, and then some prescription for the operator ordering and
the regularization must be given. These are  quite hard tasks
which have been attacked so far by many people with little success.
Further the requirement of self-adjointness on the constraint operators
in our formalism may provoke additional cumbersome problems as discussed
in \S4.4.

Here note that this requirement of self-adjointness is not
essential for the main part of our formalism because the $\VNA$ algebra $\C$
of constants of motion can be defined even if the constraint operators
are not self-adjoint and the theorems in \S3 depends only on the structure
of the $\VNA$ algebras apart from Proposition 3.10. Hence our formalism
can  be applied to the complex canonical theory based on the Ashtekar
variables\CITE{Ashtekar.A1988B,Kodama.H1993} with slight modifications.
The essential point of the argument
in \S4.4 is that the dynamics depends on the choice of the domains of
the constraint operators.

Even if a mathematically well-defined quantum totally constraint system
is constructed, there remains a problem. It is the type of the
$\VNA$-algebra $\C$, or equivalently its commutant $\W=\C'$. If it turns
out to be of type I, the formalism developed in this paper is directly
applicable. However, if it is of type II or type III, we must extend our
formalism to physical acausal subspaces which are not irreducible with
respect to the physical subalgebras.  At present we have no idea on
the type of $\C$ of quantum gravity.

In this connection we should comment on the commutation relations of
the constraint operators. In the conventional Dirac quantization of
totally constrained systems the commutation relations of the constraint
operators are regarded as having a fundamental importance because
they determine the consistency of the quantum constraint equations.
However, in our formulation, no condition is imposed on them. This is
not because they are not important but because they are implicitly
built in the algebraic structure of the relevant $\VNA$-algebras.
For example, if some commutator of two constraint operators is an
invertible operator, $\W$ coincides with all the bounded operators of the
Hilbert space and $\C$ becomes trivial.

These problems are very difficult to solve in a general form at the present
stage. However, the investigation of them in simplified theories or
subsector of the full theory such as $(2+1)$-theories, minisuperspace
models, and quantum black holes with high symmetries is possible and
is expected to give a hint on the generic situations. It is going to
be done in subsequent papers.

\section*{Acknowledgments}

Some of the main ideas of this work were conceived when I participated
in the program ``Geometry and Gravity'' in Newton Institute for
Mathematical Science.  The author would like to thank the participants
of the program, especially Jim Hartle, Karel Kuchar, Abhay Ashtekar
and Lee Smolin,  for valuable discussions and the staff of the institute
for their hospitality.  This work was supported by the Grant-In-Aid
for Scientific Research of the Ministry of Education, Science and
Culture in Japan(05640340).

\end{document}